\documentclass[12pt]{article}

\usepackage{star}
\usepackage{tikzit}
\usepackage{natbib}
\usepackage{circuitikz}
\usepackage[toc,page]{appendix}
\usepackage{longtable}
\usepackage[utf8]{inputenc}
\usepackage{authblk}

\def\scheduled{\textsf{R}_\textsf{sch}}
\def\memebership{\textsf{R}_\textsf{mem}}
\usepackage{rotating}
\usepackage{pgfplots}
\pgfplotsset{compat=1.18}
\graphicspath{{img/}{tikz/}{img/tikz/}}

\usepackage[a4paper,
            bindingoffset=0.2in,
            left=0.5in,
            right=1in,
            top=0.75in,
            bottom=0.75in,
            footskip=.25in]{geometry}

\providecommand{\keywords}[1]{\textbf{\textit{Keywords ---}} #1}

\begin{document}

\title{Pricing rides as option contracts: guarantees and memberships under travel-time uncertainty}


\author[1,2]{Mohamad Elmasri\thanks{Corresponding author: \texttt{mohamadelmasri@cunet.carleton.ca, melmasri@lyft.com}}}
\author[1]{Mingze Li\thanks{\texttt{mingzeli20000625@163.com}}}
\author[1]{Yunran Wei\thanks{\texttt{yunranwei@cunet.carleton.ca}}}

\affil[1]{Department of Mathematics and Statistics, Carleton University.}
\affil[2]{Lyft Inc.}
\date{\today}
\maketitle

\vspace{-0.25in}
\begin{abstract}
Modern ride-share platforms must commit to a price before a trip is taken, yet the realized fare depends on travel time that is uncertain at the moment of sale, which can occur days in advance. The upfront price thus decomposes into the expected fare and the premium on an insurance claim whose payoff is the shortfall between the realized and promised prices. This work computes the fair premium for such a claim, following from contingent-claim pricing. Recent advances in travel-time distribution modeling cast the asymptotic distribution of travel time as a Brownian motion on a metric graph carrying a cyclostationary travel-time process. We thus derive closed-form premiums under both a population and a route-conditional travel-time law. A single pricing equation then covers products of increasing informational difficulty, such as on-demand and scheduled rides and capped-membership and commuter products. To price on sparse data, we develop route and travel-time samplers and validate them out of sample on 2014 Quebec City GPS trips, where they reproduce the empirical trip-time distribution. We show that guarantee premiums are inexpensive, typically under 5\% of the fare. Conditioning on the realized route roughly halves the premium while reducing tail risk. Analyzing membership abuse, we show that adverse selection acts precisely along the dimension the price fails to condition on, yielding a design principle: subsidies additive in the fare, such as a fixed per-ride discount, leave marginal incentives unchanged and are abuse-proof by construction, with deterministic closed-form cost.
\end{abstract}


\section{Introduction}\label{sec:introduction}
\subsection{Background}
Mobility is vital to human activities, forming an integral component of our economic and trade networks, social interactions, political ties, and overall quality of life. Large-scale, trip-level data with extensive temporal and spatial coverage enables us to better diagnose transportation problems and develop effective solutions. Classical taxi providers priced rides based on actual time and distance traveled, with payment upon arrival. In contrast, modern ride-sharing services\footnote{Such as Waymo Inc., Lyft Inc., and Uber Inc.} must price rides well in advance. Commuter rides, for example, are priced at least a month in advance. Pricing more complex products, such as the right to travel a route for a predetermined fixed price, remains a challenge.

Two immediate challenges exist to achieve a general approach to pricing rides in transportation networks: 
understanding price variability in the marketplace and developing a modular, easily computable framework for pricing complex products. Price variability is predominantly influenced by travel uncertainty and supply and demand dynamics, the latter often controlled by ensuring sufficient driver availability. Recent progress in understanding travel uncertainty has demonstrated, both theoretically and in practical applications, that long-term deviations in travel time behave similarly to deviations from a normal distribution, despite the apparent chaotic nature of real-world traffic~\citet{elmasri2020predictive, woodard2017predicting}.

This work redefines the ride-share pricing problem as the pricing of a contingent claim, the right to a future trip at a promised price, whose value we compute with well-established financial tools. Building on the work of~\citet{elmasri2020predictive}, we price these agreements using well-established financial tools, providing a novel method to price complex transportation products, including, but not limited to, those mentioned earlier. We focus on strategically designing transportation incentives that reshape the commuter experience by offering financial guarantees that reduce pricing uncertainty\footnote{Consider a business establishment, such as a gym or medical clinic, with repeat clients who visit on regular schedule. To establish a stronger business-client relation, the business is considering providing personalized transportation packages, where clients have the option to cap their trip cost to a predetermined amount, regardless of the actual cost.}. In particular, we are interested in giving a definite answer to the following question:
\begin{quote}
    How to price a commuter membership that ensures a maximum trip cost of $K$?
\end{quote}

\subsection{Motivating example}

\begin{figure}
    \centering
    \includegraphics[width=1\textwidth]{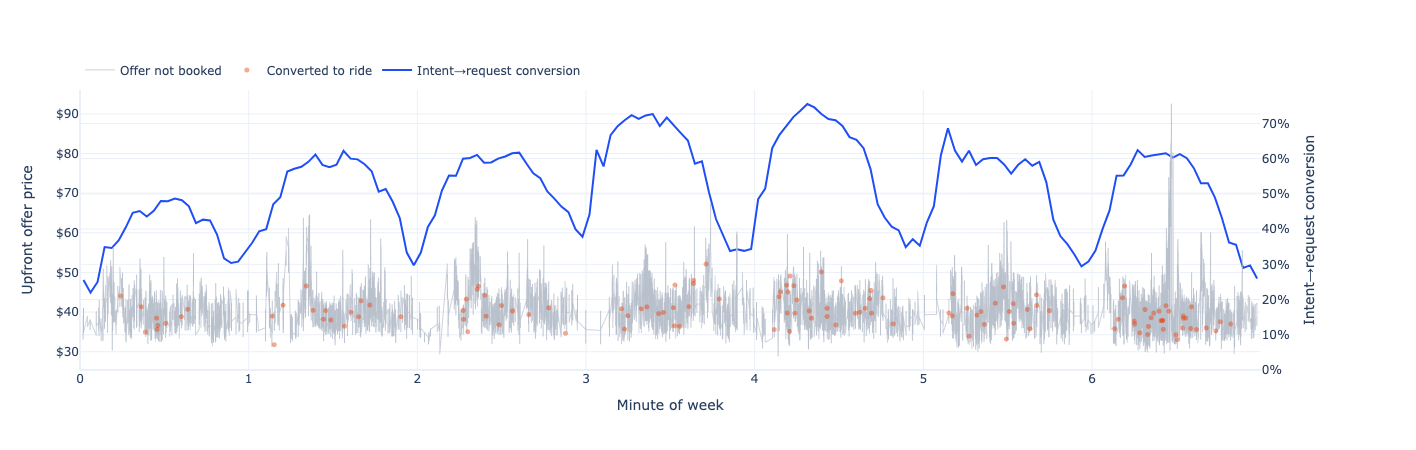}
   \caption{Offered prices (grey) and the sample of converted, paid prices
    (orange) over one week, together with the intent-to-request conversion
    rate. Location and time are anonymized for privacy.}
    \label{fig:true-price}
\end{figure}

We provide some intuition on the price distribution that is only partially
observed in a ride-share market, using real data. For a fixed route $\path$,
each departure time $t$ carries an offered price $P_\path(t)$; sweeping $t$
over a period traces out a distribution of prices rather than a single value.
Figure~\ref{fig:true-price} shows this price action over one week. Offered
prices are in grey, and the subset that converted to a paid ride is in orange.
The grey spread makes the underlying price distribution visible, while the
orange points show that we observe realized prices only where intent converted
to a request, a partial selection-shaped view of the distribution. Inference
becomes harder the less popular the origin--destination pair. To determine the upfront premium for capping a ride's price at $K$ over a period $[0,T]$, we therefore need the joint dynamics of offers and intent-to-request conversion, which are not always available.

Calculating this premium requires computing (a) the ride initiation likelihood
(conversion), (b) the probability that the uncapped price would have exceeded
$K$, and (c) the present value of any potential price exceeding $K$. Available
historical data on empirical ride initiation frequencies is statistically
sufficient to compute (a). The sparsity of ride data necessitates a theoretical
model of the price distribution to compute (b) adequately. Once (b) is
determined, (c) can be computed using known financial tools. A generalized
pricing distribution is, thus, at the core of achieving our objective of price
capping.

\subsection{Outline}
Section~\ref{sec:background} collects the necessary background: metric graphs
(Sec.~\ref{sec:metric-graph}), cyclostationary processes over them
(Sec.~\ref{sec:cyclostationary}), and the resulting travel-time process
(Sec.~\ref{sec:traveltime-process}), which we cast as an arithmetic Brownian motion
on the graph. Section~\ref{sec:assumptions} states the market assumptions under
which we price. Section~\ref{sec:pricing-main} develops the pricing equation: we
first derive (Sec.~\ref{sec:upfront}) a closed-form premium for the
on-demand price guarantee under both a population and a route-conditional
travel-time law. Later (Sec.~\ref{sec:complex-pricing}) we package this single premium into products of increasing informational difficulty, scheduled rides and memberships.

Because these premiums must be computed on sparse data, Section~\ref{sec:sampler} introduces route and travel-time samplers, and Section~\ref{sec:data} describes the 2014 Quebec City GPS data (QCD) on which we evaluate them. Section~\ref{sec:data-analysis} is the empirical core: we verify that the samplers reproduce the out-of-sample trip-time distribution (Sec.~\ref{sec:fedelity-analysis}), report the traffic-parameter estimates that feed the premium (Sec.~\ref{sec:param-estimation}), and study the guarantee's profit-and-loss behaviour as a function of the strike $K$
(Sec.~\ref{sec:strike-sensitivity}), of a bundled discount
(Sec.~\ref{sec:discount-analysis}), and of deliberate membership abuse
(Sec.~\ref{sec:selective-exercise}). Section~\ref{sec:discussion} we discuss open ended problems and analysis.

\subsection{Literature review}
\label{literature}
Previous research has shown that incorporating uncertainty into travel time estimates can lead to improved outcomes in ride-sharing systems, such as higher driver-rider match rates (up to 85\%) and lower average prices (up to 12\%) \citep{long2018ride}. These improvements can also positively impact other key system metrics, including reliability, service rate, utilization, and profitability \citep{li2021vehicle, li2022pricing,li2022ride}. Consequently, interest in understanding travel time uncertainty has been growing for practical reasons \citep{westgate2013travel,wang2019simple, hunter2009path,westgate2016large}.  Researchers have explored various approaches, ranging from models that estimate uncertainty from individual edges \citep{zheng2013urban,jenelius2013travel} to those that estimate route and travel uncertainty simultaneously \citep{westgate2013travel,wang2019simple, hunter2009path,westgate2016large}. With the increasing volume of collected GPS data, research has primarily focused on understanding the distribution of travel time through parametric methods \citep{woodard2017predicting,guo2012multistate,ma2017estimation,zhang2019novel,wu2025statistical} or by examining the limiting properties of stochastic and dynamical processes resembling real-world travel behaviour \cite{elmasri2020predictive, bolin2022gaussian}.

Recent progress has shown that deviations in travel time behave similarly to deviations from a normal distribution~\cite{elmasri2020predictive, woodard2017predicting}. Therefore, it is not surprising that regression-based models have shown promising results in travel time modelling \citep{westgate2016large, woodard2017predicting,budge2010empirical, westgate2013travel}. In particular,~\cite{elmasri2020predictive} have suggested a Gaussian form as a population-style distribution for travel time, where both the mean and variance scale with distance. For example, a \(N(n\mu, n\sigma^{2})\), where \(n\) is the number of road segments and \((\mu, \sigma^{2})\) are map- (possibly traffic time bin-) specific mean and variance constants. A similar distribution has been proposed for route-conditional travel time. 

For a survey of dynamical processes on networks (including transportation networks), see~\citet[ch. 11]{barrat2008dynamical}, for recent statistical work on the topic see~\citep{kolaczyk2014statistical, snijders2017modeling,burk2007beyond, britton2002bayesian,golightly2005bayesian, bolin2022gaussian}. 

Dynamic pricing is central to ride-sharing platforms, enabling real-time price
adjustments that balance supply and demand while maintaining service
quality~\citep{banerjee2015pricing, yan2020dynamic}. Such pricing has been shown
to increase driver availability, reduce peak-hour demand, and limit inefficient
pickups~\citep{chen2015peeking, hall2015effects}, though price volatility and
fairness remain open challenges~\citep{yan2020dynamic}. We deliberately abstract
from this two-sided, equilibrium view: our object is not the market-clearing
price but the \emph{cost} of a guarantee written on top of it, so we take the
underlying fare process as given and price contracts against it.

Pricing such a guarantee is a contingent-claim problem. The idea of valuing a
service guarantee as a financial derivative, rather than as a price to be
optimized, has precedent outside transportation, for instance \citep{rasmusson2001pricing} prices access to a virtual path across a communication network,
over a future time interval and at a strike $K$, as an option
whose underlying is a stochastic process on network resources. Our
construction is structurally analogous but instantiated on a road network. One caveat inherited from this analogy is that road travel time is not a traded, hedgeable asset, so
the market is incomplete and no replicating portfolio exists. We therefore price
the guarantee at its discounted expected cost under the modelled travel-time law
(Section~\ref{sec:assumptions}), a cost-based premium rather than a
no-arbitrage replication price.

Beyond single rides, there is growing interest in pricing more complex
transportation products, such as commuter memberships; most work here focuses on optimization methods for commuter ride-sharing~\citep{WANG2019390, MA2017345}. A membership priced at a flat cap, however, raises a selection problem familiar from flat-rate and subscription pricing: a price paid regardless of usage attracts precisely the users for whom it is least profitable and weakens their incentive to economize \citep{dellavigna2006gym, lambrecht2006paying}. The distinction between this selection channel and the usage (moral-hazard) channel is well studied in insurance \citep{powell2021disentangling}, and it is the lens we apply to membership abuse in Section~\ref{sec:selective-exercise}.

Financial instruments have been used in transportation research. \citet{Shaheen2017} organizes barriers to transportation access into a spatial, temporal, economic, physiological, and social framework, in which price uncertainty sits between the economic and temporal categories, establishing fare certainty as an access question rather than a convenience feature. \citet{fournier2023futures} propose an opt-in futures market for road tolls in which travelers prepay to lock in a price for a future window. \citet{wan2023market} supply the valuation machinery, consisting of travel time derivatives priced by risk-neutral valuation under market incompleteness. However, this system requires exchange infrastructure and a published benchmark that do not yet exist.

\section{Preliminaries}\label{sec:background}
\subsection{Metric graphs}\label{sec:metric-graph}
A finite graph $\cG = (\cV, \cE)$ consists of a set of a finite set of vertices $\cV =\cbr{v_i}$ and a set $\cE = \cbr{e_j}$ of edges connecting the vertices. We write $u \sim v$ or $\rbr{u,v} \in E$  if $u$ and $v$ are connected with an with edge. We let $\cG$ be a directed graph, that is, $\rbr{u,v}$ represents and edge from $u$ to $v$. We call a graph $\cG$ a metric graph, if every edge $e$ is assigned a length $0< l_e< \infty$, and over $e$ it possible to assign a coordinate  $c \in [0, l_e]$ which increases in a specified (but otherwise arbitrary) direction along the edge. Since the length $l_e$ are assumed to be finite, this implies that the graph is compact. A point $s \in \cG$ is a position on an edge, i.e.~$s = (e, x)$, where $x \in [0, l_e]$. A path $\path = \pathseq{\pstart}{\pend} \in \cG$ is a sequence of edges $\rbr{e_1, \dots, e_n}$ where the path starts at point $\pstart \in e_1$ and ends at $\pend \in e_n$. A natural choice of metric for the graph is the shortest path distance, which for any two points in $\cG$, it is defined as the length of the shortest path in $\cG$ connecting the two. We denote this metric by $\delta(\cdot, \cdot)$ from now on. We provide a path conditional version of this distance metric as $\delta_\path(s) = \delta_\path(s, \pstart)$, for $s \in \path$, as the distance between points $s$ and the start of the path $\pstart$, over the the path. We further let $\delta_\path$ defines the whole distance of $\path$. Let $n_\path = \abr{\cbr{e : e \in \path}}$ be the number of edges of $\path$. We assume that the graph is connected, so that there is a path between all vertices. We define $\Pi(e_1, e_2)$ to be the set of all paths between edge $e_1$ and $e_2$, and similarly $\Pi(\pstart, \pend)$ be the set of all paths between points $\pstart, \pend \in \cG$. 
There is no unique way of defining process over graphs. This is a result of various way of defining the boundary condition (vertex condition), such as the Neumann-Kirchhoﬀ condition, see~\citet[Eq~(1.4.4) and Them~1.4.4]{berkolaiko2013introduction}. In this work, we treat a path $\path = \pathseq{\pstart, e_1}{e_n,\pend}$, as an extended sequence of an uninterpreted edge, having the geometry
\[\sbr{0,\;  \delta(\pstart, e_1) + \sum_{e\in \path} l_e + \delta(e_n, \pend)}, \]
without introducing intermediate vertices at $l_e,$ $e \in \path$.

For every edge $e\in \cE$, a function $f_e \in L_2(e)$ if $f_e$ is square-integrable over the edge $e$. We further define a function $f$ over $\cG$ as a collection of functions $f = \cbr{f_e}_{e \in cE} \in L_2(\cG)$ if $f_e \in L_2(e)$ for each $e \in \cE$. This results in $L_2(\cG)$ being the direct sum over $L_2(e)$ as 
\[L_2(\cG) = \bigoplus_{e \in \cE} L_2(e). \]

We finally let $C(\cG) = \cbr{f \in L_2(\cG): f \text{ is continuous}}$ be the space of continuous functions on $\cG$. We define $\RR_+$ to be the positive part of the real line.

\subsection{Cyclostationary processes over metric graphs}\label{sec:cyclostationary}
We let $X=X(t, s)$, for every fixed time index $t \in \RR_+$ and point $s \in \cG$, be a 2nd order cyclostationary process in the wide sense. Here $X(t,s)$ is spatio-temporal process defined as
\begin{equation} \label{eq:speed}
X(t,s) = m(t, s) + \epsilon(t, s), \qquad t \in \RR_+, s \in \cG,
\end{equation}
where, in the temporal domain (holding the spatial fixed), for every $e \in \cE$ and $t, u \in \RR_+$, we have
\begin{subequations}
\label{eq:mean-covariance}
\begin{align}
    m(t, s) & =\EE[X(t,s)] = m(t + T_e, s), \label{eq:mean}\\
    \sigma^2(t, s) &= \EE[\epsilon^2(t,s)] = \sigma^2(t +T_e, s), \label{eq:variance}\\
    \varsigma(t, u, s) &= \EE[\epsilon(t,s)\epsilon(u, s)] = \varsigma(t +T_e, u+T_e, s) \label{eq:covariance},
\end{align}
\end{subequations}
where $T_e \in \RR_+$ is an absolute positive constant defining the temporal cycle of edge $e$. Here, $m(t,s), \sigma^2(t,s), \varsigma(t,u, s) \in C(\cG)$ (measurable functions over $\cG)$, and $\EE[\epsilon(t,s)]=0$ for all pairs $(t, s) \in \RR_+ \times \cG$.  We further let $\max_{e\in \cE} T_e \leq T_\cG < \infty$ be the global temporal cycle over $\cG$. That is, all equalities in~\eqref{eq:mean-covariance} apply when replacing $T_e$ by $T_\cG$. By definition of cyclostationarity, for come absolute constant $U_1>0$, we have
\[\abr{m(t,s)} + \abr{\sigma^2(t,s)} \leq \max_{u \in [0, T_\cG]}\cbr{m(u, s), \sigma^2(u, s)} \leq U_1,\]
for all $t\in \RR_+$, which is a stronger condition than the classical growth condition imposed on stochastic random variables, see~\citet[Thm.~5.2.1]{oksendal2013stochastic}. We further assume Lipschitz continuity in space and time as 
\begin{equation}\label{eq:lip-condition}
    \abr{m(t,x) - m(u, y)} + \abr{\sigma^2(t,x) - \sigma^2(u,y)} \leq C_2 \cbr{\abr{t-u} + \abr{x-y}},
\end{equation}
for $t, u \in \sbr{0, T_\cG}$, $x,y \in \cG$, and $C_2>0$ is an absolute constant.~\eqref{eq:lip-condition} results in spatial Lipschitz continuity when $t=u$, and temporal when $x=y$. 

We assume a conditional independence condition between space and time, that is 
\[X(t, x) \perp X(t, y) \;\given\; t; \qquad x,y \in \cG, x\neq y.\] 

We define a ride on $\cG$ to be the pair $(\rho, t)$, representing the path $\rho$ of the ride and the start time $t_2>0$.

\subsection{Travel time process}\label{sec:traveltime-process}
We first define a transportation network $\cN = (\cG, X)$ to be a metric graph $\cG$ equipped with wide-sense cyclostationary process $X(t, s)$, $t \in \RR_+$ and $s \in \cG$. Travel time over a path $\path = \pathseq{\pstart}{\pend} \in \cG$ can be directly computed as the stochastic integral 
\begin{equation}
    \label{eq:traveltime-process}
\T_\path = \int_\path X(\T_s, s)ds = \int_\path m(\T_s, s)ds + \int_{\path} \epsilon(\T_s,s)ds, \quad \T_\pstart = 0
\end{equation}
where $\int_\path$ represent integration over the predetermined path $\path$, with $s \in \path$ being a point over the path. Here, $\T_s, s \in \path$ defines the travel time up to location $s$ on $\path$, such that the travel time at the start point $\pstart$, $\T_\pstart =0$. ~\cite{elmasri2020predictive} approximated the integral~\eqref{eq:traveltime-process} by a discrete version $\T_\path^d$, as 
\[\T^d_\path = \sum_{e \in \path} m_e(\tau_e) + \epsilon_e(\tau_e), \]
where 
\begin{equation}\label{eq:marginals}
    m_e(t) = \int_e m(t,s)ds, \quad \sigma^2_e(t) = \int_e \sigma^2(t,s)ds, \quad \epsilon_e(t) = \int_e \epsilon(t,s)ds,
\end{equation}
for $t \in \RR_+$.~\eqref{eq:marginals} are the spatial marginals over $X(t,s)$ defined in~\eqref{eq:speed}. We have implicitly assumed that all edges are traversed in $\T_\path^d$, and that $\T_\path^d$ is defined on a transportation network $\cN^d$ defined by the marginals of $X$, such that $\cN^d = (\cG, X^d)$, and $X^d = \cbr{X_e(t)}_{e \in \cE}$, where $X_e(t) = \int_e X(t,s)ds$. The following lemma is an adaptation of \citet[Lem~6, Them.~6]{elmasri2020predictive} that characterize the asymptotic distribution of $\T_\path^d$. 

\begin{lemma}[Population distribution~\cite{elmasri2020predictive}] \label{lem:population-distribution}Consider $\path$ as a random walk over the $\cN^d$ and let $\tau = \cbr{\tau_e}_{e \in \path}$ be the random arrival times over edges of the walk $\path$. Let  $\tau$ be sufficiently mixing\footnote{at least $\rho$-mixing, see~\cite{bradley2005} and \citet[Lem.~5,6]{elmasri2020predictive}}. Then as the number of edges in $\path$ grow to infinity (i.e.~$n_\path \to \infty)$, we have
\begin{subequations}\label{eq:population-asymp-dist}
\begin{align}
    n_\path^{-1}\T^d_\path &\stackrel{\text{a.s.}}{\to} \mu \\
    n_\path^{-1/2}(\T^d_\path - n_\path\mu) &\stackrel{d}{\to} N(0,\sigma^2), \label{eq:population-distribution}
\end{align}
\end{subequations}
where $\mu$ and $\sigma^2$ are absolute positive constants that are independent of initial conditions and of $\path$. 
\end{lemma}

The result in~\eqref{eq:population-asymp-dist} suggest that deviations of $\T_\path$ around the average travel time per edge $\mu$, normalized by $\sqrt{n_\path}$, resemble independent samples from a normal distribution, with variance that is independent of the route. However,~\eqref{eq:population-asymp-dist} is an asymptotic distribution over a population of rides in a system. That is for an arbitrary ride from a population of rides, the probability it arrives before $n_\path\mu$ is $1/2$. For a specific ride, say $(\rho_i, t_2)$, this probability can be significantly different if $\T_{\path_i}$ has not reached the asymptotic distribution. The following lemma is an adaptation of~\citet[Thm.~7]{elmasri2020predictive} that characterized the distribution of travel time for a specific ride. First, we define a deterministic mean and covariance sequences for a ride over $\path=\pathseq{e_1}{e_{n_\path}}$, starting at $t_1\in \RR_+$, as
\begin{subequations}\label{eq:mean-covariance-seq}
    \begin{align}
        t_{i+1} &= t_{i} + m_{e_i}(t_i), \quad i \in \cbr{2,\dots, n_\path},\\
        \mu_\path &= \sum_{i =1}^{n_\path} m_{e_i}(t_i), \\
        \sigma^2_\path &= \sum_{i=1}^{n_\path} \sigma^2_{e_i}(t_i) + 2 \xi \sum_{i=2}^{n_\path}\sigma_{e_i}(t_i)\sigma_{e_{i-1}}(t_{i-1}),
    \end{align}
\end{subequations}
for some absolute constant $\xi$. Here, $m_e$ and $\sigma^2_e$ are the spatial marginals defined in~\eqref{eq:marginals}. 

\begin{lemma}[Route-predictive distribution, Thm.~6 in~\cite{elmasri2020predictive}] \label{lem:route-predictive} Under the conditions of Lemma~\ref{lem:population-distribution}, for an arbitrary start time $t_0 \in \RR_+$, given a path $\path=\pathseq{e_1}{e_{n_\path}} \in \cG$, we have 
\begin{equation}\label{eq:route-distribution}
    \sigma^{-1}_{\path} \rbr{\T^d_\path - \mu_\path} \stackrel{d}{\to} N(0, \eta), \quad \text{as  } n_\path \to \infty,
\end{equation}
where $\eta$ is a strictly positive constant that is independent from initial conditions, and $\sigma^2_\path $, $\mu_\path$ are defined in~\eqref{eq:mean-covariance-seq}. 
\end{lemma}

The result of Lemma~\ref{lem:route-predictive}, although still suggests approximate normality of travel time distribution, it integrates route-specific information in the mean and covariance functions~\eqref{eq:mean-covariance-seq} to better approximate this distribution. As shown in~\citet[Fig.~3]{elmasri2020predictive}, the approximation in~\eqref{eq:route-distribution} can achieve significant accuracy when compared to~\eqref{eq:population-distribution}. 

Working with $\T_\path$ in~\eqref{eq:traveltime-process} directly is challenging, since the distribution of $\epsilon$ is not well understood. Every edge $e \in \path$ can influence $\T_\path$ differently, depending on the cyclostationarity pattern of the variance $\sigma(t,s)$ and the actual error distribution. 

To overcome such challenges, we rely on the results in Lemma~\ref{lem:population-distribution} and~\ref{lem:route-predictive}. The mean and covariance functionals $\mu(t,s)$, $\sigma(t,s)$ in~\eqref{eq:mean-covariance} define manifolds over the graph $\cG$. Using the asymptotic normality of travel and those functionals, we redefine travel time as the stochastic process, initiated at $\pstart=0$, as
\begin{equation}\label{eq:stochastic-travel-time}
    d\T_{s} = \alpha(\T_{s}, s)ds+ \beta(\T_{s}, s)dB_s,\qquad \T_{\pstart} = 0,   \quad s \in \path,
\end{equation}
where $B_s$ is Brownian process.~\eqref{eq:stochastic-travel-time} suggests that travel time over a route $\path$ behaves like an arithmetic Brownian motion on the spatial domain, having the expectation and variance as 
\[\EE[\T_s] = \int_\path \alpha(\T_s,s)ds \quad \Var[\T_s]  = \int_\path \beta^2(\T_s, s)ds, \quad s \in \path,\]
where the variance follows by the assumption that $B^2(t,s)$ is bounded, and thus by the Itô isometry property \citet[Lem.~3.1.5]{oksendal2013stochastic}, we have $\EE\rbr{\int_\path \beta(\T_s, s)dB_s}^2 = \int_\path \beta^2(\T_s,s)ds$. 

Finally, to adapt $\T_\path$ to the population version in Lemma~\ref{lem:population-distribution} and the route-specific version in Lemma~\ref{lem:route-predictive}, we let $\alpha$ and $\beta$ be 

\begin{equation}\label{eq:mean-variance-sde}
    \alpha(t,s) = 
    \begin{cases}
        \mu & \text{population} \\
        m(t,s) & \text{route-specific}
    \end{cases}
    \qquad 
    \beta(t,s) = 
    \begin{cases}
        \sigma & \text{population}        \\
        \sqrt{\eta}\sigma(t,s) & \text{route-specific},
    \end{cases}
\end{equation}
where $m(t,s)$ and $\sigma(t,s)$ are as defined in~\eqref{eq:mean-covariance} and $\eta$ is a universal parameter that is estimated from data and accounts for residual error.

\section{Market setting, information, and pricing assumptions}\label{sec:assumptions}
The pricing literature commonly distinguishes cost-based, competition-based, and value-based approaches \cite{phillips2021pricing}. We consider a single-provider, partial-equilibrium setting. The underlying metered fare is taken as exogenous and is determined by the tariff in \eqref{eq:price}. Our object is not to optimize the market-clearing fare, but to calculate the expected provider cost of a guarantee written on top of that fare. The monopoly assumption allows the provider to implement the tariff and guarantee premium without modeling competitor responses; it does not imply that the resulting price is profit-maximizing.

The analysis begins when a ride request is submitted and ends when the ride is completed. Matching, waiting-time determination, driver repositioning, and supply-demand equilibrium are outside the model. We impose the following assumptions.

\begin{enumerate}[label={(A$_{\arabic*}$)}, ref={(A$_{\arabic*}$)}]  
	\label{monopoly}
	\item Completion conditional on acceptance. Once a ride request is accepted, the rider does not cancel and the trip is completed without rematching or interruption. The trip is not assumed to have zero duration.
	
	\item Monopoly pricing environment. The platform is modeled as a monopolist and can implement the underlying tariff and guarantee premium without accounting for competitive responses. This assumption isolates the guarantee-cost calculation from competition; it does not imply that the resulting price is profit-maximizing, since demand and competitor behavior are not optimized within the model.
	
	\item Driver availability. A driver is available for every accepted request at the relevant pickup time $t_0$. The matching process and any endogenous pickup delay are not modeled.
	
	\item Exogenous baseline usage. In the baseline pricing model, ride initiation, utilization, and route choice are treated as exogenous to the guarantee. Riders and drivers do not change their behavior in response to the guarantee premium. Section \ref{sec:selective-exercise} partially relaxes this assumption by allowing a member to select among otherwise eligible rides using a private signal correlated with the realized fare. The model does not include additional ride demand induced by the membership.
	
	\item Product-specific information. Information at pricing depends on the product. For an on-demand ride ($P_1$), the route and pickup time are known. For a scheduled ride ($P_2$), the origin, destination, and scheduled pickup time are known, while the realized route and travel-time shock remain uncertain. For an unrestricted membership ($P_3$), the contract terms, horizon, ride limit, and historical distributions are known at subscription, but the individual future rides are not. For the commuter membership in Section \ref{sec:commuter}, the origin, destination, and departure-time window are additionally known at subscription, while future route choices, traffic shocks, and realized travel times remain unknown.
\end{enumerate}

We further impose the following financial conventions.
\begin{enumerate}[label={(B$_{\arabic*}$)}, ref={(B$_{\arabic*}$)}]  
	\item Discounting. The risk-free rate $r$ is constant and known, and all time quantities are converted to the same units as $r$. The empirical analysis sets $r=0$ because the relevant horizons are short. \label{frictionless}

	\item   Frictions and premium principle. Taxes, transaction costs, administrative expenses, capital charges, and additional platform fees are excluded. The baseline premium is the discounted expected provider payout under the modeled travel-time law; it contains no additional risk loading and is not a replication price. Unless otherwise stated, all distributions and expectations are understood relative to the information available at the relevant pricing time, and this conditioning is suppressed from the notation.
	
	\end{enumerate}
Under our assumptions, riders are price-insensitive but may increase usage in response to external conditions (i.e, abuse  as mentioned in section \ref{sec:commuter}).  This setup reflects a monopoly market with non-strategic supply, where the platform acts as the sole decision-maker, and the customer accepts the offered price without negotiation or cancellation. The pricing method is not explicitly adjusted based on real-time fluctuations in driver supply or rider demand. 
	
Our objective is to propose a cost-based pricing method and evaluate its stability and volatility under dynamic conditions, such as varying traffic or membership abuse, without modeling equilibrium outcomes in a two-sided market. 	We explicitly exclude the matching process, strategic driver behavior, and real-time supply-demand equilibrium mechanisms from our analysis.

While dynamic pricing based on real-time demand and supply can, in theory, maximize revenue, it is often operationally costly and computationally intensive. In practice, rider behavior is highly unpredictable and sensitive to wait time and price fluctuations, leading to instability. The tradeoff in our method  is a potential loss in revenue in exchange for greater pricing stability and predictability.

\section{Pricing equation in ride-share markets}\label{sec:pricing-main}
Ride pricing is determined by three constant factors predefined by the service provider. A fixed price $C_0$ (starting price), a cost for a unit of time $C_1$, and a cost for a unit of distance $C_2$. Let $P_\path(s, t)$ be the price of a ride that starts at time $t>0$, over the route $\path$, and at location $s \in \path$, that is
\begin{equation}\label{eq:price}
    P_\path(s, t) = C_0 + C_1\T_s(t) + C_2 \delta_\path(s), \qquad s \in \path, 
\end{equation}
where $\T_s(t) = \T_{\path(s)}(t)$ is the travel time up to point $s \in \path$ and starting at time $t>0$ and $\delta_\path(s) = \delta(s,\pstart)$ is the length of the path $\path$ up to point $s$. To simplify notations, we will drop the $(t)$ from $\T_s(t)$, since the start time defined in the pricing function $P_\path(s, t)$. The base price is determined at the start point $\pstart \in \path$, that is 
\[P_{\pstart}(\pstart, t) = C_0, \quad \text{for all } t > 0, \]
and the final price is determined at the destination point $\pend$, and arrival time $t_2$, as
\[P_\path(\pend, t) = C_0 + C_1\rbr{t_2 - t} + C_2\delta_\path.\]

Here,  $\T_{\path(\pend)} = t_2 -t$, an actual realization. Otherwise, for any point $s \in \path \setminus \cbr{\pstart, \pend}$, the price equation~\eqref{eq:price} is a random variable determined by $\T_s(t)$. Moreover, for any two points $s_1, s_2 \in \path$, if $s_2$ is further along $\path$ than $s_2$ (i.e.~$\delta_\path(s_1) \leq \delta_\path(s_2)$), then 
\[P_\path(s_2,t) \geq P_\path(s_1,t).\]

Such inequality in the time domain is not true, that is $P_\path(s, t_1)$ is not necessary less than $P_\path(s,t_0)$ for $t_1 < t_0$, and $s \in \path$. The price equation~\eqref{eq:price} is, thus, composed of three components: (i) the fixed price $P_\path(0,0) = C_0$, (ii) the incremental price $C_2\delta_\path(s)$, and (iii) the variable price $C_1\T_s, s \in \path$. In this sense, the true price of a ride over $\rbr{t,\path}$ that arrived at $\T_\path = t_2 - t$, is 
 \begin{equation}\label{eq:true-price}
    P_\path(\pend, t) = C_0 + C_1(t_{0} -t) + C_2 \delta_\path,
 \end{equation}

 If the realization is not yet observed, then the present value of the ride price at start location $\pstart$ is a random variable having the expectation
\begin{equation}\label{eq:predicted-price}
    P_\path(\pstart, t) = \EE\sbr{P_\path(\pend, t)}= C_0 + C_1\EE\sbr{\T_\path} + C_2 \delta_\path,
 \end{equation}

In~\eqref{eq:predicted-price}, the expectation is taken over all possible realization of $\T_\path$, without accounting for the value of money. For example, having the option to allocate the cost of the ride, say $K$, to an investment vehicle with a risk-free annualized interest rate $i$, would earn $Ke^{i\T_\path}$ by the time the ride arrives. In this sense, the price of a ride $(t, \path)$ at the start location $\pstart$ is 
\[P_\path(\pstart, t) = \EE\sbr{e^{-i\T_\path} P_\path(\pend, t)}. \]

Adjusting for the value of money is mostly mathematical exercise, since in most real-world applications, arrival times are within minutes. Exceptions exist, for example cargo and freight, which can last weeks. Such adjustment can be useful for schedules of rides or membership pricing as shown in later sections.

Finally, the present value of the pricing formula can be extended to route unconditional pricing through expectations. Given the set of all path $\Pi\rbr{\pstart, \pend}$  between origin-destination pair $(\pstart, \pend)$. The price of the ride  $(t, \pstart, \pend)$ is
\[P_{\rbr{\pstart, \pend}}(\pstart, t) = \EE\sbr{e^{-i\T_{\path_k}}P_{\path_k}(\pend, t)},\qquad \path_k \sim \Pi(\pstart, \pend),\]
where the expectation is taken over the uniform distribution of routes, and travel time. A discrete probability measure can be applied on the set $\Pi$, for a non-uniform distribution.

\subsection{Up-front pricing} \label{sec:upfront}
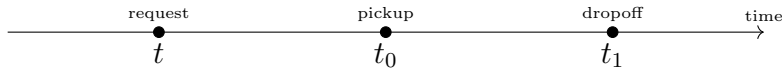
\begin{figure}[th!]
\centering
\begin{tikzpicture}
  \draw[->] (0,0) -- (10,0) node[above]{\tiny time}; 
  \filldraw[black] (2,0) circle (2pt) node[below] {$t$} node[above] {\tiny request};
  \filldraw[black] (5,0) circle (2pt) node[below] {$t_0$} node[above] {\tiny pickup};
  \filldraw[black] (8,0) circle (2pt) node[below] {$t_1$} node[above] {\tiny dropoff};
\end{tikzpicture}
\caption{Time progression of a ride request, pickup and dropoff.}
\label{fig:ride-time}
\end{figure}

Up-front pricing, a popular initiative by ride-share providers, promises a price $K$ at the request time $t$ for a ride over the path $\path$ starting at location and time $(\pstart, t_0)$, $t_0\geq t$, and potentially ending a location and time $(\pend, t_1)$. When $t_0=t$, we call this an on-demand ride, otherwise this is a scheduled ride. The progression if a ride request is shown in Figure~\ref{fig:ride-time}. 

The upfront price $K$ is guaranteed unless the realized price $P_\path(\pend, t_1)$ does not exceed a cutoff threshold, which is often defined as a percentage of the upfront price, say $Ke^\zeta$, for a predefined constant $\zeta \in \RR_+$ predefined. Under Assumption~\ref{frictionless}, a request if accepted is always fulfilled. let $R(t_1, t_0, \path, K)$ be the cost of this liability at the time of execution (i.e.~the arrival time $t_1$), defined as 
\begin{equation}\label{eq:ride-cost}
    R(t_1, t_0, \path, K) = \rbr{P_\path(\pend, t_1) - K}^+\one_{\cbr{P_\path(\pend, t_1) \geq Ke^\zeta}}. 
\end{equation}

In this sense, the final price charged to the customer at time $t_1$ is
\begin{equation}\label{eq:charged-upfront-price}
K + R(t_1, t_0, \path, K) = 
\begin{cases} 
P_\path(\pend, t_1) & \text{if } Ke^\zeta \leq P_\path(\pend, t_1) \\
K & \text{otherwise}.
\end{cases}
\end{equation}
It is clear from~\eqref{eq:ride-cost} that for the promised ride price $K_1 \geq K_2$, we have 
\[R(t_1, t_0, \path, K_1) \leq R(t_1, t_0,, \path, K_2).\]

Oftentimes rideshare providers do not explicitly estimate the cost of such a guarantee at time the request time $t$, that is $R(t, t_0, \path, K)$. Instead, consumers are only charged $K\$$ at time $t_0$. At arrival time $t_1$, the price is adjusted according to~\eqref{eq:charged-upfront-price}.  

Under Assumptions listed in Section \ref{sec:assumptions}, we model the travel time along the path $\path = \pathseq{\pstart}{\pend}$ as an asymptotic stochastic dynamic governed by the partial differential equation~\eqref{eq:stochastic-travel-time}. The price of the guarantee~\eqref{eq:ride-cost} at request time $t$ is then given by:
\begin{equation}\label{eq:european-knock-in}
\begin{aligned}
 R(t,t_0, \rho, K) &= \EE\sbr{e^{-r(t_0-t)}R(t_1,t_0, \path, K)} \\
  &=e^{-r(t_0-t)} e^{-rd_4}\sbr{\cbr{d_0 - rC_1d_1^2- K}\cbr{1-\Phi\rbr{d_3}} + C_1d_1 \phi(d_3)},
    \end{aligned}
\end{equation}
where 
\begin{align*}
    d_0 &= C_0 + C_1\EE[\T_\path] + C_2\delta_\path, &
    d_1 &= \cbr{\Var[\T_\path]}^{1/2}, &
    d_2 &= \frac{Ke^\zeta -d_0}{C_1d_1}, \\
    d_3 &= d_2 + rd_1, &
    d_4 &= \EE[\T_\path] - rd_1^2/2. &&
\end{align*}
The intermediate steps are carried out in Appendix~\ref{app:derivations}.

When $r=0$, we have $d_3 = d_2$ and the solution to \eqref{eq:european-knock-in} simplifies to 
\[ R(t,t_0, \rho, K) = (d_0-K)\cbr{1- \Phi(d_2)} + C_1d_1\phi(d_2).\]

Further if $\zeta=0$, there is no minimum threshold for the guarantee and the same expression
applies with $d_2 = \rbr{K - d_0}/\rbr{C_1d_1}$.

The next section introduces more complex pricing structures with unknown routes and start times, and packages them into membership offering. 

\subsection{Pricing complex products}\label{sec:complex-pricing}
Pricing of on-demand single rides in Section~\ref{sec:pricing-main} is done under the simplest case of known (route, start time)$=(\path, t_0)$. We vary these two quantities by generalizing from a known fixed route to only the (start, end) $=\rbr{\pstart, \pend}$ locations are known, and then an unknown start time. Summarized in three pricing products
\begin{enumerate}[label={(P$_{\arabic*})$}, ref={(P$_{\arabic*}$)}] 
    \item On-demand. An instantaneous ride request to go from  $\pstart$ to $\pend$ in $\cG$ over the route $\path$. \label{on-demand}
    \item Scheduled. A ride request to go from $\pstart$ to $\pend$ in $\cG$ at a future time $t$. \label{scheduled}
    \item Membership. An option to request up to $M$ rides within a specified time interval, $T_M \in \RR_+$, at a predetermined price $K$. Such rides can be (but not necessarily) conditioned over fixed hours and locations.  \label{membership}
\end{enumerate}

Computing an accurate upfront price--paid at the time of request or subscription--is the central challenge in three scenarios of increasing complexity:~\ref{on-demand}, \ref{scheduled}, and \ref{membership}. The information available at the pricing time diminishes with each case. In the on-demand scenario \ref{on-demand}, both the route's start/end points and the real-time network dynamics are known. For a scheduled ride~\ref{scheduled}, only the start/end locations are known, requiring a forecast of network dynamics. Finally, under a membership plan~\ref{membership}, no details are known at the pricing stage, and the ride itself is elective. 

The premium price of the on-demand~\ref{on-demand} scenario is defined in~\eqref{eq:european-knock-in}. Recall that $\Pi(\pstart, \pend)$ is the set of routes between any two points, $\pstart$ and $\pend$. Let $\pi(\path)$ be the likelihood of route $\path$ in $\Pi(\pstart, \pend)$. If a uniform distribution is desired, then $\pi(\path) \propto 1/\abr{\Pi(\pstart, \pend)}$. $\pi(\path)$ can (not necessarily) time dependent. The premium price of the scheduled ride~\ref{scheduled} requested time $t$ for (start time, start location, end location) = $(t_0, \pstart, \pend)$ is
\begin{equation}
\begin{aligned}\label{eq:scheduled-price}
    \scheduled(t, t_0, K) & = 
\sum_{\path_i \in \Pi\rbr{\pstart, \pend}} R(t, t_0, \path_i, K)\pi(\path_i) \\
& = e^{-r(t_0-t)} \sum_{\path_i \in \Pi\rbr{\pstart, \pend}} \EE\sbr{e^{-r\T_{\path_i}}R(t_1, t_0, \path, K)}\pi(\path_i),
\end{aligned}
\end{equation}
where the conditioning step is deferred to Appendix~\ref{app:derivations}.

The pricing in~\eqref{eq:scheduled-price} occurs by averaging over all routes between $\pstart$ and $\pend$, hence the notation $\bar R$. In the same spirit, let $\pi(t)$ be the distribution of pickup times on the interval $T_R$. The price guarantee of a membership~\ref{membership} purchased at time $t$ of $M$ rides on the interval ride~\ref{scheduled}, for fixed price $K$, is
\begin{equation}\label{eq:membership-price}
    \memebership(t, M) = M \sum_{t_0 \in T_R} e^{-r\rbr{t_0 - t}} \scheduled\rbr{t_0, t_0, K}   \pi(t_0),
\end{equation}
where $\scheduled$ follows from~\eqref{eq:scheduled-price} and the derivation from the double sum over pickup times and routes is given in Appendix~\ref{app:derivations}. To validate our method, the next section develops a statistical sampling methods for routes and travel time. 

\section{Statistical sampling of travel times and routes}\label{sec:sampler}
To simulate alternative routes to a realized trip, we develop a statistical approach for trip sampling, and demonstrate its statistical fidelity. Given a trip (route, start time)$=(\path, t_2)$, we alternative routes, then we sample travel times.
\subsection{Route sampling}\label{sec:route sampling}
When all traffic states of the graph is observed, routing-finding algorithm, such as the one proposed by~\cite{1970algorithm}, can be used to list all routes from two points ($e_1, e_n)$ on the $\cG$ at the desired start time $t_2$. Marrying this route sampling approach to reality is intuitive, but not always possible. Existing traffic data do not cover all states of the graph.  Our objective is propose a sample method that provide statistically equivalent routes, that are graph-invariant for all triplet (start location, end location, start time) = ($e_1, e_n, t_2)$. 

Given a route $\langle e_1, e_2 \dots, e_n \rangle$, first, we match edges $\times$ time bins using a mean matching method. or each edge $e \in E$ and time bin $b \in B$, let $n_{e,b}$ be the sample size used to calculate the sample mean $\hatm_e(t_b)$ and variance $\hatsigma_e(t_b)$. We define statistically equivalent edges by the difference between their means, as 
\begin{equation}\label{eq:similar-edges}
E_{e, b}(\zeta) =
\cbr{
    \rbr{e', b'} \in E \times B:
    {\frac
    {\abr{\hatm_e\rbr{b} - \hatm_{e'}\rbr{b'}}}
    {\sqrt{\frac
        {\hatsigma_e(b)}
        {n_{e,b}} 
        +
        \frac
        {\hatsigma_{e'}(b')}
        {n_{e',b'} }
    }} \leq \zeta
    }
    } 
\end{equation}

The set in~\eqref{eq:similar-edges} is constructed using simultaneous t-tests, with unequal variance and sample size, on the hypothesis that the two edges $e, e'$ have equivalent means under the time bins $b, b'$. The critical value $\zeta$ can be chosen to achieve a $(1-\beta)100\%$, $\beta \in (0, 1)$ level of significance of the Student's t distribution with  degrees of freedom following the Welch–Satterthwaite equation~\citep{satterthwaite1946approximate,welch1947generalization}, as
\[
\textsf{degrees of freedom} = 
\frac{
\rbr{{\frac
        {\hatsigma_e(b)}
        {n_{e,b}} +
    \frac
        {\hatsigma_{e'}(b')}
        {n_{e',b'} }
    }}^2
}{
 \frac{\rbr{\hatsigma_e(b)/ n_{e,b}}^2}{ n_{e,b}-1} 
 + 
  \frac{\rbr{\hatsigma_{e'}(b')/ n_{e',b'}}^2}{ n_{e',b'}-1}
}
\]

Second, we randomize the length of the sampled route using a normal distribution centered at the number of edges $n$ of a given realized route. This results in a route sampling process as
\begin{equation} \label{eq:route_sampling}
\begin{aligned}
    t_0 & \sim \textsf{given start time} \\
    \path = \langle e_1, e_2 \dots, e_n \rangle & \sim \textsf{given route } \path \\
    E_{\path} &= \cbr{E_{e, b}: (e,b) \in \path }\\
    n'& \sim {\lfloor}\textsf{Normal}\sbr{n, \tau}{\rfloor}\\
    \path' = \langle e'_1, e_2 \dots, e'_{n'} \rangle & \sim \textsf{Uniform}\sbr{E_\path}. 
\end{aligned}
\end{equation}

Here $ {\lfloor} Z {\rfloor}$ represent the integer floor of a positive variable $Z$.   

\subsection{Travel time sampling}\label{sec:travel time sampling}
We define $B$ to be the set of time bins of the data, for example, EveningRush, WeekDay, MorningRush, EveningNight, and WeekendDay, used in~\cite{elmasri2020predictive}. For every edge $e \in \cG$, let $\hatm_e(t_b)$ and $\hatsd_e(t_b)$ be the sample average speed and standard deviation of duration to travel $\T_e$, respectively. Here $t_b \in B$, meaning, i.e.~a time bin. We reserve the notation $t_b$ to represent the time bin $t$ falls into. This results in $\abr{E} \times \abr{B}$ number of traffic states in $\cG$. 

Following the simple auto-regressive structure and the exact sampler proposed by~\cite{falk1999simple}, we sample route-specific travel time $\T_\path$ as 
\begin{equation} \label{eq:travel_time_sampling}
\begin{aligned}
    t_0 & \sim \textsf{given start time} \\
    \langle e_1, e_2 \dots, e_n \rangle & \sim \textsf{given route } \path \\
    \xi & \in \rbr{-1,1} \textsf{ a correlation coefficient} \\
    \rbr{\Sigma}_{ij} & = \xi^{\abr{i-j}}, 1 \leq i, j \leq n\\
    (Z_1, \dots, Z_n)  & \sim \textsf{multivariate-Normal}\sbr{0, 2\sin\rbr{\pi \Sigma/6}} \\
    (U_1, U_1, \dots, U_{n}) & = \rbr{\Phi\cbr{Z_1}, \dots, \Phi\cbr{Z_n}} \\
     \T_{e_i} &\sim \textsf{log-Normal}\sbr{\hatm_e\rbr{t_{b_i}},\hatsd_e\rbr{t_{b_i}}; \rbr{U}_i} \\
     t_{b_{i+1}}& = t_{b_{i-1}} + \T_{e_i} \\
    \T_\path & = \sum_{i=1}^{n_\path} \T_{e_i}. 
\end{aligned}
\end{equation}
The sinusoidal transformation of $\Sigma$ ensures its positive definiteness, otherwise the nearest positive definite correlation matrix can be used~\cite{higham2002computing}. The sample mean and standard deviations are used, since the population version is rarely accessible in real-world applications.

The sampling method~\eqref{eq:travel_time_sampling}, is iterative. The travel times for the first edge at the start-time traffic-bin is sampled, and iteratively, the travel time of the second edge at the traffic bin of the start-time plus the travel time of the first edge is sampled, and so on until the last edge. This insures an $\alpha$-mixing fully-dependent. A consequence of the constructed dependence in the uniform series of random variables $(U_1, \dots, U_n)$. A non-iterative, route-invariant, population version of this sampling process is
\begin{equation} \label{eq:travel_time_sampling-population}
\begin{aligned}
    t_0 & \sim \textsf{given start time} \\
    \langle e_1, e_2 \dots, e_{n_\path} \rangle & \sim \textsf{given route } \path \\
    \T_\path & = n_\path \hatmu.
\end{aligned}
\end{equation}

When the population mean is not accessible, we replace it by the sample mean $\hatmu$. For more details how to estimate $\hatmu$ refer to \cite[Sec.~3.3.2]{elmasri2020predictive}.

\section{Quebec City trip data} \label{sec:data}
\subsection{Data collection}
Quebec City 2014 GPS data (QCD) is collected using the \textit{Mon Trajet} smartphone application developed by Brisk Synergies Inc. This study uses a sample of open data,\footnote{{A cleaned sample of the data is available at \texttt{https://github.com/melmasri/traveltimeCLT}}.} which contained 21,872 individual trips. The sample contains no data that can be linked to individual drivers. While no data was collected during the winter months, the precise duration of the collection period is kept confidential. {The application was installed voluntarily by over 4,000 drivers, who then anonymously logged information using a simple interface. The exact number of drivers is kept confidential.}
\subsection{Data cleaning}
No measure was provided to ensure the validity of trips, i.e. if they were made solely by motor vehicles and not walkers or cyclists, and excluded non-traffic interruptions such as parking. Data is thus cleaned by breaking down long trips if there was more than 4 minutes of idle time or over 2 minutes between consecutive GPS observations. The start and end points of trips were trimmed. A trip officially began when the vehicle speed first exceeded 10 km/h and ended when it last dropped below 10 km/h. Trips were excluded if their median speed was less than 20 km/h, their maximum speed was less than 35 km/h, or their total driving distance (calculated from GPS points) was less than 1 km. After cleaning, the dataset contained 19,967 trips. 

The median trip duration was 19 minutes (average 21 minutes), with the longest trip being 3 hours and 27 minutes. The median trip distance was 14.5 km (average 16.6 km), with a maximum of 170.4 km. The median time between consecutive GPS points was 4 seconds (average 9 seconds). Refer to \cite[Sec.6.3] {elmasri2020predictive} for more details on the cleaning process. 

\subsection{Map matching}
A third-party service (TraxMatching\footnote{{\texttt{https://www.traxmatching.io}}}) was used to map trips' GPS observations to the Quebec City road network mapped by The OpenStreetMap Project\footnote{\texttt{https://www.openstreetmap.org}} (OSM), a publicly accessible open-source project. This process is called map-matching, and numerous high-quality methods are available to do this \citep{newson2009hidden,hunter2013large}. For each trip, the third-party service returns a sequence of mapped GPS points with lengths equal to the original sequence. Each mapped GPS point is associated with a source ``node id'', ``way id'' and destination ``node id'' corresponding to a unique directional edge whose ``way id'' is between the source and destination nodes. The map-matching process resulted in 46,386 unique directional edges, which constitute the travelled portion of Quebec City, {not its entirety.} The average edge length is 170~meters, and the median is 81~m. 

\subsection{Traffic estimation}
We estimate the total travel time per edge by calculating i) within-edge travel time, as the time spent within the edge, and ii) across-edge travel time, as the time spent between the two closest GPS observations, where one is in the edge and the other is in an adjacent edge. We calculate the across-edge distance in the same way. The total travel time per edge is then 100\% of within-edge plus across-edge travel time, weighted by half the proportion of across-edge distance to the total length of the edge. Total edge lengths are obtained from OSM. In rare circumstances, the map-matching service also returns intermediate edges that do not have initial GPS observations. This happens, for example, when a vehicle is moving fast or through a tunnel. We treat those intermediate edges, those without GPS observations, as a single edge and calculate the total travel time over it, and then assign edge travel time proportionally to the length of each intermediate edge. With these total travel time estimates, we calculate the average speed per edge by dividing the total travel time by the total length.

\begin{figure}[!htbp]
  \centering
  \includegraphics[width=0.6\textwidth]{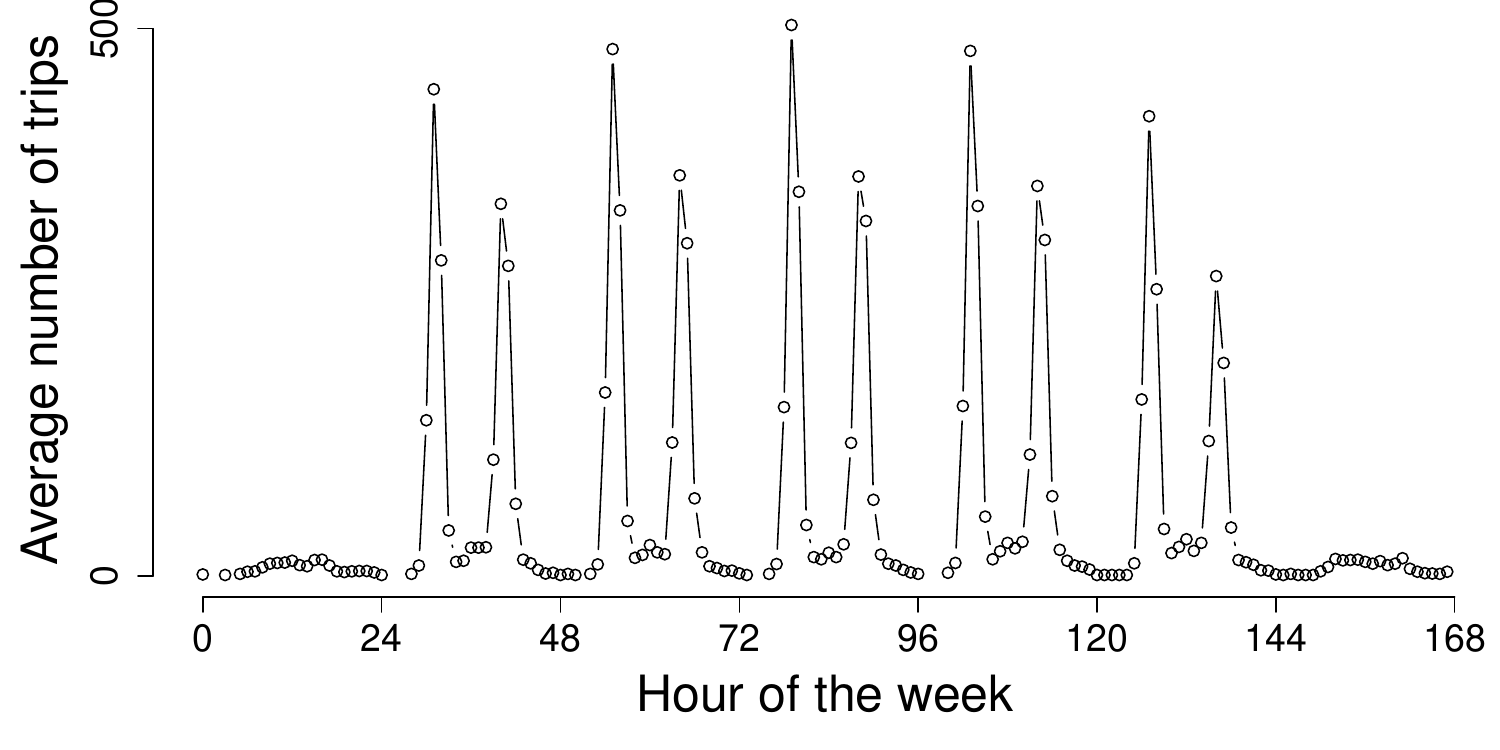}
  \caption{Average number of trips per hour in QCD, by hour of the week.}
  \label{fig:quebec-city-time-per-hour}
\end{figure}

Figure \ref{fig:quebec-city-time-per-hour} shows seasonal (weekly) traffic patterns per week hour, starting at the first hour of Sunday. The volume of traffic is reduced overnight on weekdays starting after 7~p.m.~and on weekends. Daily traffic peaks are associated with a.m.~and p.m.~rush hours, with strong dips in between.
\subsection{Data splitting} \label{subsec:split}
Because of the seasonal (weekly) traffic pattern illustrated in Figure~\ref{fig:quebec-city-time-per-hour} and the sparsity of GPS data for each edge, we classify our traffic data into three traffic time-bins (traffic-bins): i) an ``AM-rush hour'' bin for weekdays 6:30–8:30AM; ii) a ``PM-rush hour'' bin for weekdays 3:30–5:00~p.m., and iii) a ``Non-rush hour'' bin for all remaining periods. Alternative traffic-bins have been tested, but we found that aggregating data into these bins yielded the best results. In QCD, 37\% of trips occurred in an AM-rush hour, with the {same proportion} for the PM-rush hour, and 26\% in all other periods.  

A test set of $2{,}000$ trips is drawn uniformly at random without replacement, leaving the remaining $21{,}054$ trips as training data. We classify a trip into a bin if all trip-edges are travelled within that bin; otherwise, the speed data for those trips are used for traffic estimation, but not trip analysis. Because the draw is unstratified, an edge$\times$traffic-bin state observed only in a held-out trip has no training observation; such states fall back to the pooled ``Global'' bin at prediction time.

\begin{figure}[h!tbp]
    \centering
    \includegraphics[width=\linewidth]{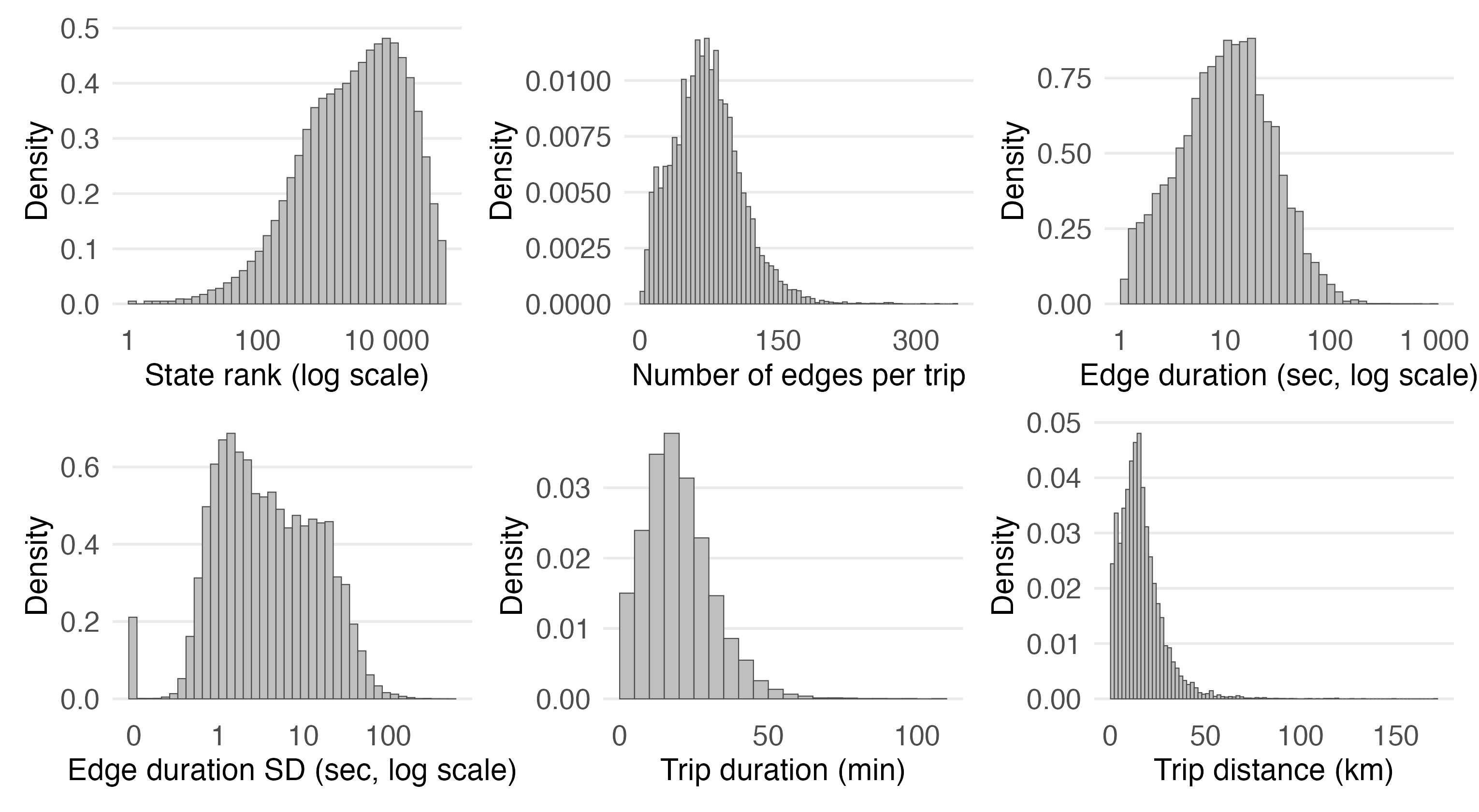}
    \caption{Descriptive statistics of the edge $\times$ time-bin representation in QCD.
    (Row~1, left) Empirical visit likelihood over state rank on a log scale (edge $\times$ time-bin states
    sorted in decreasing visit frequency), shown as a density histogram: a small fraction
    of states accounts for most observed traffic,
    motivating the frequency-weighted edge sampling in~\eqref{eq:route_sampling}.
    (Row~1, centre) Density of the number of edges per trip, supporting the
    route-length randomization $\mathrm{Normal}[n,\tau]$ in~\eqref{eq:route_sampling}
    and the large-$n_\path$ regime of
    Lemmas~\ref{lem:population-distribution}--\ref{lem:route-predictive}.
    (Row~1, right) Visit-weighted density of per-state mean edge travel time on a log scale (sec).
    (Row~2, left) Visit-weighted density of per-state SD of edge travel time on a log scale (sec).
    (Row~2, centre) Density of trip travel times.
    (Row~2, right) Density of trip distances.}
    \label{fig:edge-timebin-descriptive}
\end{figure}

\section{Data analysis}\label{sec:data-analysis}
 
\subsection{Outline}
We evaluate the pricing framework of Sections~\ref{sec:pricing-main}--\ref{sec:complex-pricing}
on the Quebec City data (QCD, Section~\ref{sec:data}) along the three products of increasing
informational difficulty introduced in Section~\ref{sec:complex-pricing}: on-demand
rides~\ref{on-demand} (route and start time known), scheduled rides~\ref{scheduled} (only
origin, destination, and start time known), and memberships~\ref{membership} (no ride
information at subscription). We first establish the fidelity of the route and travel time samplers of
Section~\ref{sec:sampler} reproduce the empirical trip distribution
(Section~\ref{sec:fedelity-analysis}), then report the traffic-parameter estimates that feed the
pricing formula, and finally study the guarantee premium~\eqref{eq:european-knock-in} and its
profit-and-loss behaviour as a function of the strike price $K$, of a bundled discount, and of
deliberate membership abuse.
 
Throughout, prices use the QCD tariff $C_0 = 3.17$~CAD, $C_1 = 0.31$~CAD/min and
$C_2 = 0.9$~CAD/km, the risk-free rate is set to $r=0$, and, unless stated otherwise, the
strike is set to the expected price at the start of the ride,
$K = \hat\EE\sbr{P_\path(\pstart, t_0)}$. The expectation $\EE\sbr{\T_\path}$ and variance
$\Var\sbr{\T_\path}$ entering~\eqref{eq:european-knock-in} are estimated by the two models of
Section~\ref{sec:traveltime-process}: a \emph{trip-specific} model that conditions on the
route through the edge-level mean and variance functionals (Lemma~\ref{lem:route-predictive})
and a route-invariant \emph{population} model (Lemma~\ref{lem:population-distribution}).

\subsection{Fidelity of the sampling process}\label{sec:fedelity-analysis}
To justify pricing on simulated rather than observed rides, we first verify that the samplers
of Section~\ref{sec:sampler} reproduce the empirical trip-time distribution out of sample. We
hold out 2{,}000 test trips, fit the edge\,$\times$\,time-bin statistics on the remaining
training trips, and, for each test trip, draw $1{,}000$ travel times from each sampler and
average them into a single trip estimate; the population sampler~\eqref{eq:travel_time_sampling-population}
requires no averaging and is drawn once per trip. We compare the full-order dependent
sampler~\eqref{eq:travel_time_sampling} against three reduced-dependence variants
(independent, $\xi=0$; first-order; second-order), the population sampler, and the
two-state Hidden Markov model of~\cite[Algo.~2]{woodard2017predicting}; the correlation
parameter is fixed at $\xi = 0.31$~\cite[Tab.~1]{elmasri2020predictive}.
 \begin{figure}[h!tbp]
    \centering
    \includegraphics[width=0.45\linewidth]{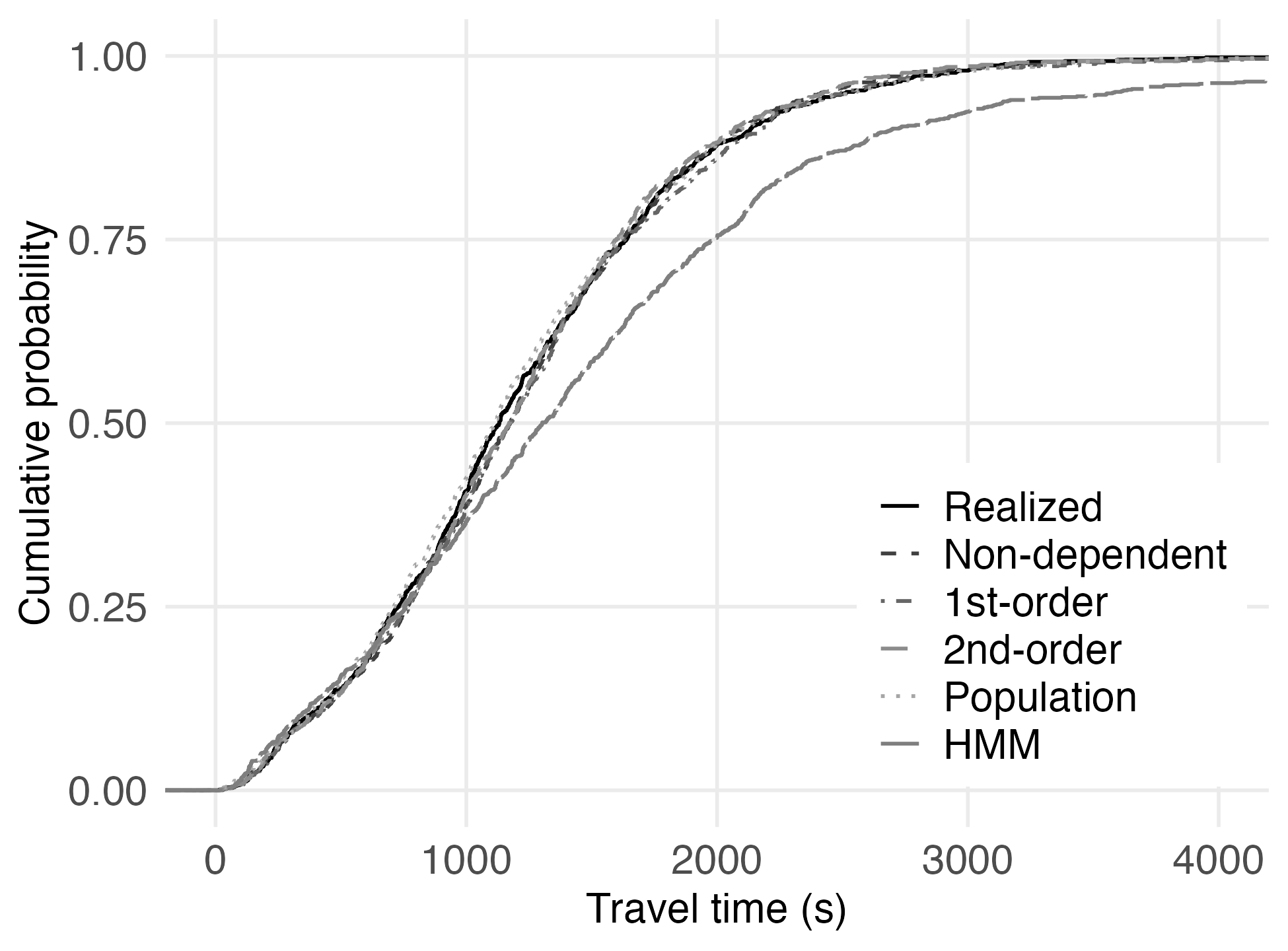}
    \includegraphics[width=0.45\linewidth]{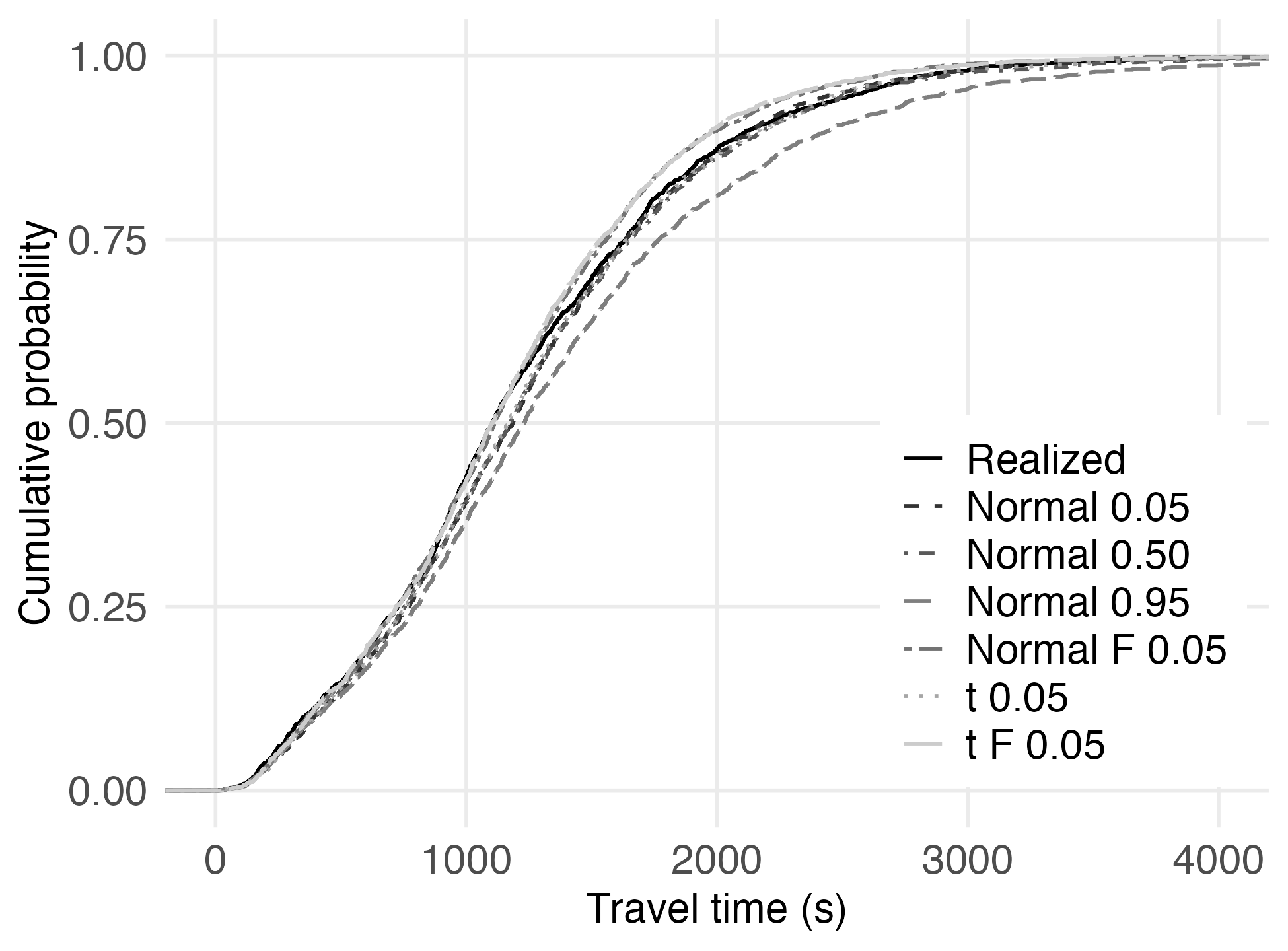}
    \caption{(Left) Comparing the empirical cumulative distribution of the 2000 test trips, our proposed travel time sampling methods, and the HMM model of~\cite[Algo.~2]{woodard2017predicting}. (Right) Route-varied travel time simulation via~\eqref{eq:route_sampling} under normal and $t$-based edge matching at significance levels $0.05$, $0.50$, and $0.95$.}
    \label{fig:fidelity_traveltime}
\end{figure}
 
Figure~\ref{fig:fidelity_traveltime}(left) shows that our samplers track the empirical
cumulative distribution closely. All correlation-accounting variants are nearly
indistinguishable from one another, and the marginal gain from higher-order dependence beyond
the independent draw is small. Most route-level uncertainty is already captured by the
edge\,$\times$\,time-bin marginals together with the log-normal edge model. The population
sampler is visibly more dispersed. Every one of our samplers improves substantially on the HMM
alternative, which systematically overstates trip duration across the support.

Route variation~\eqref{eq:route_sampling} preserves this fidelity: resampling each test trip
into five statistically equivalent routes (matched by the simultaneous $t$-tests
of~\eqref{eq:similar-edges}) and averaging their sampled travel times reproduces the empirical
curve under normal edge matching at the $5\%$ significance level
(Figure~\ref{fig:fidelity_traveltime}, right); loosening the similarity threshold degrades the
match, as expected.

The descriptive statistics in Figure~\ref{fig:edge-timebin-descriptive} explain why the procedure is well posed on QCD: a
small fraction of edge\,$\times$\,time-bin states accounts for most observed traffic
(motivating the frequency-weighted edge sampling), the per-trip edge counts are large enough to
support the asymptotics of Lemmas~\ref{lem:population-distribution}--\ref{lem:route-predictive},
and the per-state mean log travel time is roughly flat while its standard deviation grows only
for the rarer, less-sampled states.

\subsection{Profit-and-loss accounting}\label{sec:pnl}
For a ride priced at strike $K$ with collected premium $R = R(t,t_0,\path,K)$ from
\eqref{eq:european-knock-in} and realized price $P = P_\path(\pend,t_1)$, the provider's
per-ride profit under cutoff $\zeta$ is
\begin{equation}\label{eq:profit}
\textsf{profit} \;=\; R \;-\; \rbr{P - K}^{+}\,\one_{\cbr{P \ge K e^{\zeta}}},
\end{equation}
i.e.\ the premium kept minus the payout triggered only when the realized price exceeds the
guarantee threshold. We summarize a population of rides with five quantities: the share of
profitable rides (\emph{win rate}), the \emph{profit factor} (total gains divided by the
absolute total of losses, so values above one indicate net profitability), the \emph{average
profit} per ride, the \emph{maximum single-ride loss}, and the average premium expressed as a
percentage of the trip price, $R/P$. The premium ratio is the headline quantity for the rider,
the profit factor and maximum loss for the provider.

\subsection{Traffic-parameter estimation}\label{sec:param-estimation}
The population model needs only the pair $\cbr{\mu,\sigma}$; we use the unbiased sample mean
$\hatmu$ and the profile estimator $\sigmaprofile$ of~\cite[Sec.~3.3.2]{elmasri2020predictive},
computed from $1{,}000$ trips. Unlike $\hatmu$, which is stable across subsamples,
$\sigmaprofile$ is not: it varies by a factor of nearly eight across random draws of
$500$ trips from the same corpus, for the reason set out beneath
Table~\ref{tb:parameter-estimates}. The
trip-specific model additionally needs the edge-level pairs $\cbr{\hatm_e,\hatsigma_e}$, the
correlation $\xi$ and the residual variance $\nu$ of Lemma~\ref{lem:route-predictive},
estimated from $4{,}000$ trips. Table~\ref{tb:parameter-estimates} reports all estimates, both
pooled (``at random'') and stratified by traffic regime. Rush-hour travel is both slower and
markedly less predictable than off-peak travel: the mean edge duration rises from
$13.70$ to $17.34$~s and its standard deviation from $18.44$ to $26.91$~s, while the residual
variance roughly doubles ($1.15$ vs.\ $2.04$). The edge-to-edge correlation is stable across
regimes ($\hatxi \approx 0.29$--$0.31$), consistent with the value used in the sampler.

\begin{table}[!htp]
  \centering
    \caption{Parameter estimation under different sampling methods}
    \label{tb:parameter-estimates}
    {\footnotesize 
  \begin{tabular}{l lllll}
                      & \multicolumn{3}{c}{Sampling method}                                \\
                      & At random         & \multicolumn{2}{c}{Stratified by traffic-bins} \\[5pt]
                     
                      &                   & {rush} & {Non-rush}                         \\
    \(\hatmu\)        & 16.46             & 17.34     & 13.70                              \\   
    \(\sigmaprofile\) & 25.18             & 26.91     & 18.44                              \\
    \midrule
    \(\hatxi\)        & 0.30              & 0.31     & 0.29                               \\
    \(\hatnu\)        & 1.74              & 2.04      & 1.15
  \end{tabular}
  }
  \par\vspace{4pt}

\end{table}

\subsection{Sensitivity to the strike price $K$}\label{sec:strike-sensitivity}
We price the on-demand guarantee~\ref{on-demand} for $10,000$ trips with $r=0$ and $\zeta=0$, so that~\eqref{eq:european-knock-in} reduces to~\eqref{eq:price-when-r=0}, and sweep the strike from the expected price ($\times 1$) to a $5\%$ and $10\%$ markup ($\times e^{0.05}, \times e^{0.1}$). Table~\ref{tab:strike-metrics} reports the outcome. 

\begin{equation}\label{eq:price-when-r=0}
R(t_0,t_0, \rho, K) = (d_0-K)\cbr{1- \Phi(d_2)} + C_1d_1\phi(d_2),
\end{equation} 
where, as in Section~\ref{sec:upfront},
\[
d_0 = C_0 + C_1\EE[\T_\path] + C_2\delta_\path, \qquad
d_1 = \cbr{\Var[\T_\path]}^{1/2}, \qquad
d_2 = \frac{Ke^\zeta -d_0}{C_1d_1}.
\]

Raising the strike makes the guarantee less likely to trigger, so the win rate climbs monotonically, from $0.75$ to $0.92$ for the population model and from $0.67$ to $0.95$ for the trip-specific model. The average premium charged to the rider falls correspondingly. The premium is modest throughout, between $0.4\%$ and $4.4\%$ of the trip price. At every strike the trip-specific premium is roughly half the population premium (for instance $2.15\%$ vs.\ $4.43\%$ at $\times 1$), a direct consequence of the tighter variance obtained by conditioning on the route. Both models are profitable on average (profit factor above one at all strikes), but the population model carries the larger tail risk, with a worst-case single-ride loss near $\$29$ against $\$22$ for the trip-specific model.

\begin{figure}
    \centering
    \includegraphics[width=0.32\textwidth]{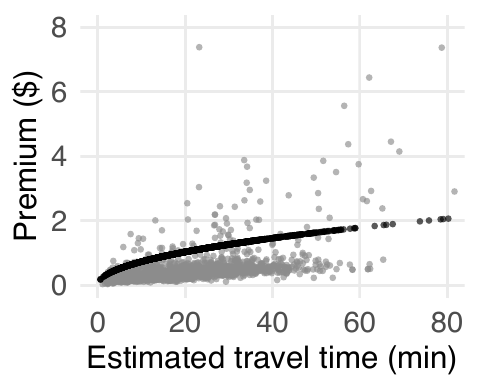}
    \includegraphics[width=0.32\textwidth]{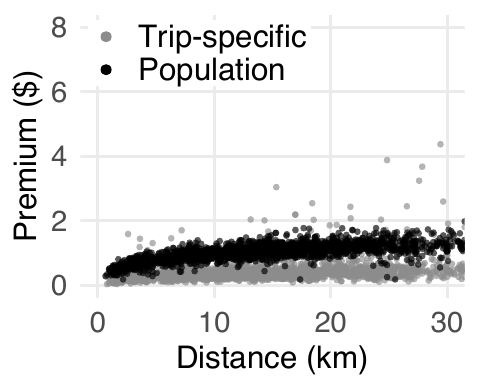}
    \includegraphics[width=0.32\textwidth]{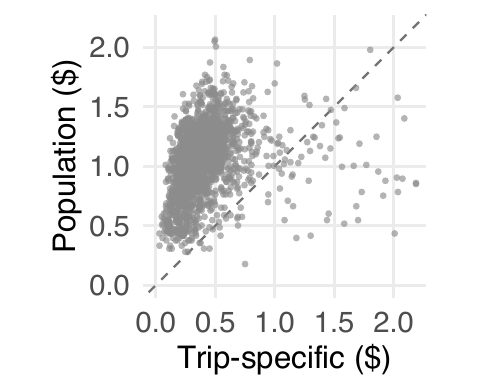}
    \caption{Premiums of the on-demand guarantee~\ref{on-demand} over $2{,}000$ held-out QCD
    test trips at $K = P_\path(\pstart, t_0)$ and $\zeta = 0$: (left) premium $R$ against
    estimated travel time; (middle) premium against trip distance; (right) population
    premium against trip-specific premium (dashed line: $y = x$). Trip-specific and
    population models are distinguished by grey and black dots, with a single
    legend shown in the middle panel.}
    \label{fig:route-specific-R}
\end{figure}

\begin{figure}[h!]
\centering
\includegraphics[width=0.42\textwidth]{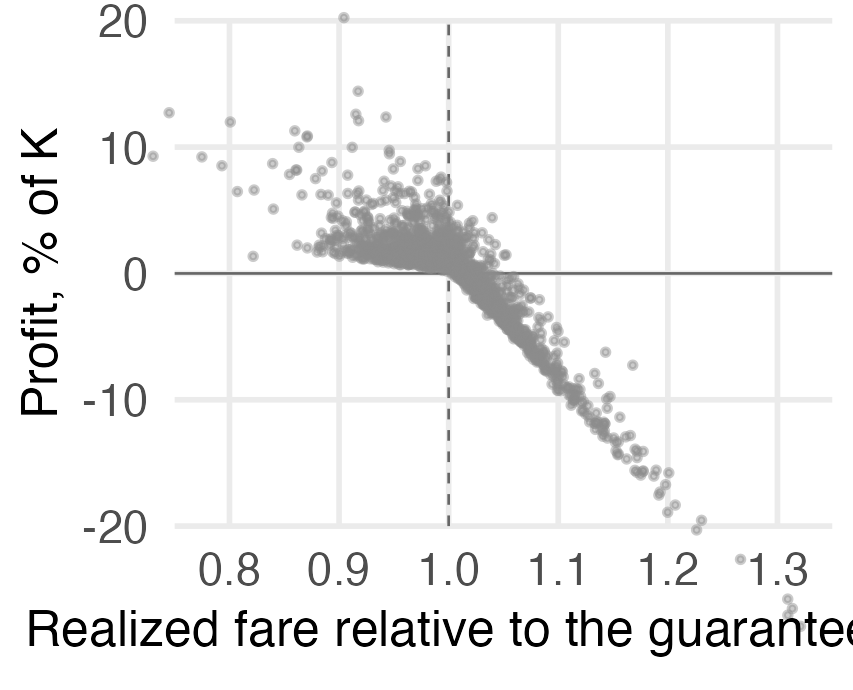}
\hfill
\begin{tikzpicture}
  \def\rOverK{0.035} 
  \pgfmathsetmacro{\rPct}{100*\rOverK}
  \pgfmathsetmacro{\kPlusR}{1+\rOverK}
  \begin{axis}[
    width=0.42\textwidth,
    height=0.34\textwidth,
    xmin=0.75, xmax=1.35,
    ymin=-20, ymax=20,
    xlabel={Realized fare relative to the guarantee, $P/K$},
    ylabel={Profit, \% of $K$},
    axis lines=left,
    axis line style={thick, gray!70!black},
    tick align=outside,
    tick style={gray!70!black},
    xtick={0.8,0.9,1.1,1.2,1.3},
    ytick={-20,-10,0,10,20},
    grid=both,
    grid style={line width=0.2pt, gray!20},
    clip=false,
    label style={font=\fontfamily{cmss}\selectfont\normalsize},
    tick label style={font=\fontfamily{cmss}\selectfont\small},
    every axis plot/.append style={line cap=round},
  ]
    \addplot[gray!45, dashed, line width=0.6pt] coordinates {(1,-20) (1,20)};
    \addplot[gray!45, dashed, line width=0.6pt] coordinates {(0.75,0) (1.35,0)};

    \addplot[draw=none, fill=gray!12]
      coordinates {(0.75,0) (1,0) (1,\rPct) (0.75,\rPct)} -- cycle;
    \addplot[black, line width=0.9pt, domain=0.75:1, samples=2] {\rPct};

    \addplot[draw=none, fill=gray!6]
      coordinates {(1,0) (\kPlusR,0) (1,\rPct)} -- cycle;
    \addplot[black, line width=0.9pt, domain=1:\kPlusR, samples=2]
      {100*(1+\rOverK - x)};

    \addplot[black, line width=0.9pt, dashed, domain=\kPlusR:1.35, samples=2]
      {100*(1+\rOverK - x)};

    \addplot[only marks, mark=*, mark size=1.3pt, black]
      coordinates {(1,\rPct) (\kPlusR,0)};
    \node[anchor=north, font=\fontfamily{cmss}\selectfont\small, inner sep=0pt]
      at (axis cs:1,-22) {$K$};
    \node[anchor=north west, font=\fontfamily{cmss}\selectfont\small, inner sep=0pt]
      at (axis cs:\kPlusR,-22) {$K{+}R$};
    \node[anchor=east, font=\fontfamily{cmss}\selectfont\small, inner sep=2pt]
      at (axis cs:0.75,\rPct) {$R$};
  \end{axis}
\end{tikzpicture}
\caption{Realized and theoretical profit of the on-demand guarantee. (Left) Per-ride profit
against the realized fare, each scaled by that trip's own guarantee $K$, over the $2{,}000$
held-out trips priced with the trip-specific model at $K = P_\path(\pstart,t_0)$, $r=0$ and
$\zeta=0$; realized fares use the observed edge durations, so nothing on this panel is
simulated except the premium. (Right) The theoretical profile on the same scale: the premium
$R$ is retained whenever the realized fare stays below $K$, profit then declines linearly and
breaks even at $K+R$ (illustrated at $R/K \approx 3.5\%$).}
\label{fig:profit_loss}
\end{figure}
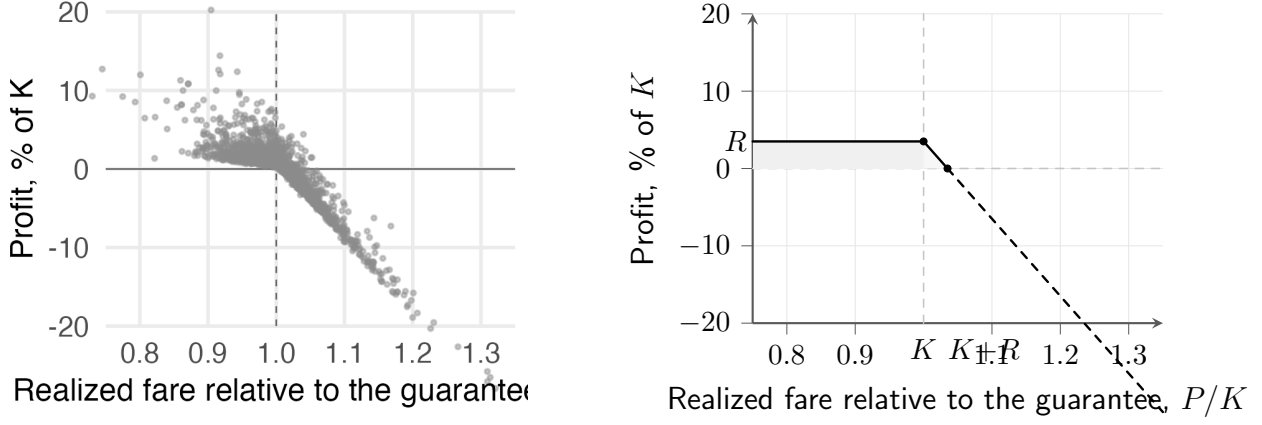

 \begin{table}[h!]
    \centering
    \caption{On-demand price-guarantee performance under different strike prices,
    with $r=0$ and $\zeta=0$, over 10{,}000 trips;
    profit factor is the ratio of total gains to total losses.}
    \begin{tabular}{l l c c c c c}
    \hline
    Model & Strike & \% profitable & \shortstack{Avg.\ profit\\(\$)} & \shortstack{Max.\ loss\\(\$)} &
    \shortstack{Profit\\factor} & Premium/$P$ (\%) \\
    \hline
    Population    & $\times 1$        & 0.75 & 0.18 & $-28.79$ & 1.39 & 4.43 \\
                  & $\times e^{0.05}$ & 0.84 & 0.11 & $-27.58$ & 1.41 & 2.35 \\
                  & $\times e^{0.1}$  & 0.92 & 0.04 & $-25.88$ & 1.36 & 1.11 \\
    Trip-specific & $\times 1$        & 0.67 & 0.04 & $-21.85$ & 1.12 & 2.15 \\
                  & $\times e^{0.05}$ & 0.87 & 0.03 & $-20.73$ & 1.23 & 0.73 \\
                  & $\times e^{0.1}$  & 0.95 & 0.05 & $-19.36$ & 1.99 & 0.36 \\
    \hline
    \end{tabular}
    \label{tab:strike-metrics}
\end{table}
These magnitudes translate into very inexpensive guarantees in absolute terms. Consistent with
Figure~\ref{fig:route-specific-R}, the premium grows essentially linearly in distance and
travel time, and for a typical $4$~km ride, priced at
$K = 3.17 + 0.31\times(0.0982\times 4000/60) + 0.9\times 4 \approx \$8.8$. The trip-specific guarantee costs on the order of \$0.2--\$0.5. Knowing the realized route therefore prices the guarantee at roughly half the cost of the population approach, with materially smaller downside. The profit profile in Figure~\ref{fig:profit_loss} matches this picture: maximum profit $R$ is retained whenever the realized price stays below $K$, profit then declines linearly to the break-even point $K+R$, and the loss beyond that point is bounded because real network trips have finite duration. The left panel traces that profile out of sample. Roughly two thirds of the held-out trips settle at the full premium, the realized fare stays within about $\pm 20\%$ of the guarantee, and the worst ride of the $2{,}000$ loses $\$7.95$, so the tail is short in both directions.

\subsection{Discount memberships}\label{sec:discount-analysis}

We next bundle a flat $10\%$ discount into the guarantee, fixing the strike at a constant
fraction of the expected price,
\begin{equation}\label{eq:discount-strike}
    K = 0.9 P_\path(\pstart, t_1)
\end{equation}
in the guarantee~\eqref{eq:ride-cost}, and vary the cutoff threshold $\zeta \in \cbr{0, 0.1}$
separately. The product we have in mind is a $40$-ride monthly membership. No ride is
known at subscription, so we mimic the distribution of potential rides: for each day
$y = 1, \dots, 30$ we draw the number of rides $M_y$, and for each ride a traffic bin $b_k$ and a route $\path_k$, as
\begin{equation} \label{eq:counterfactual_rider}
\begin{aligned}
    M_y     & \sim \textsf{Binomial}(2, 2/3), \\
    b_k     & \sim \textsf{Uniform}\sbr{\textsf{AM-rush, PM-rush, Otherwise}}, \quad k = 1, \dots, M_y, \\
    \path_k & \sim \textsf{Uniform over rides in bin } b_k \textsf{ in the training set}.
\end{aligned}
\end{equation}

The Binomial mean leads to $30 \times 2 \times \tfrac{2}{3} = 40$ expected rides per month; the membership premium is the sum of the per-ride premiums averaged over this distribution.

Because both the premium and the accumulated profit are sums over rides, the effect of the
discount is already visible one ride at a time. 
Table~\ref{tab:discount-metrics} evaluates~\eqref{eq:profit} over $100$ sampled trips priced at
the discounted strike~\eqref{eq:discount-strike}, and a monthly figure follows by scaling with
the $40$ expected rides. Its percentage columns divide each ride by its \emph{own} realized
fare before averaging, so the third column is the mean of
$\rbr{R_i - \rbr{P_i - K_i}^{+}}/P_i$ and the last is the mean of $R_i/P_i$; neither is a ratio
of aggregates. The trip-specific $-0.45\%$ therefore sits alongside a mean loss of $\$0.23$ on
fares averaging $\$24.47$, the gap between the two reflecting that the larger shortfalls fall
on the more expensive rides. Over a $40$-ride month that shortfall accumulates to roughly
$\$9$.

The table shows that the discount reverses the
ranking observed in the on-demand case. The population model stays slightly profitable, with an average return of $2.7$--$2.9\%$, whereas the thinner trip-specific premium no longer covers the discounted guarantee and turns marginally loss-making, at $-0.4$ to $-0.5\%$. 
Introducing a non-zero cutoff $\zeta = 0.1$ lowers the premium charged to the rider but also erodes the provider's margin and deepens the worst-case loss, since fewer overruns are recovered. 
This suggests that the cost-based premium has little slack once a fixed discount is accounted for.  
The route conditioning that makes the on-demand guarantee cheap also leaves the discounted trip-specific product without a buffer against realized overruns.

 \begin{table}[h!]
    \centering
    \caption{Per-ride performance of the discounted guarantee $K = 0.9\,P$ under different
    cutoff thresholds $\zeta$, over 100 trips. Both percentage columns are averages of
    per-ride ratios.}
    \begin{tabular}{l l c c c c}
    \hline
    Model & Cutoff $\zeta$ & Profit/$P$ (\%) & Avg.\ profit (\$) & Max.\ loss (\$) & Premium/$P$ (\%) \\
    \hline
    Population    & $0$   & $2.86$  & $0.09$  & $-8.41$ & 11.42 \\
                  & $0.1$ & $2.71$  & $-0.03$ & $-9.28$ &  9.73 \\
    Trip-specific & $0$   & $-0.45$ & $-0.23$ & $-6.90$ & 10.49 \\
                  & $0.1$ & $-0.53$ & $-0.36$ & $-7.99$ &  7.90 \\
    \hline
    \end{tabular}
    \label{tab:discount-metrics}
\end{table}

\subsection{Commuter memberships}\label{sec:commuter}
Consider a monthly membership~\ref{membership} of $M = 30$ rides between a
fixed origin and destination, departing at a fixed hour of the day, with the price of \emph{each}
ride capped at a promised $K$ that is quoted once and held for the whole month. Origin, destination, and departure time are known at subscription.
Traffic and route uncertainty remain unknown at time of subscription. 

We construct a counterfactual commuter from an origin--destination pair of QCD trips holding at
least $h+1$ trips, where $h$ is the number of trips that are already observed. One trip on the pair is \emph{held out} and the remaining $h$ form the commuter's
\emph{history}. A departure time is drawn from the pair's trips and every trip on both sides is
shifted to it, so that the whole month is priced and realized at the same hour and traffic bin.
The month's $M$ rides are $M$ route randomizations~\eqref{eq:route_sampling} of the held-out
trip, each carrying an independently sampled travel
time~\eqref{eq:travel_time_sampling}; the membership itself is priced
from~\eqref{eq:membership-price} with $r = 0$, with $\pi(t_0)$ degenerate at the commuter's
departure time. Route sampling is applied to the history $h$ 10 times, where the strike and premium is averaged across those samples. Since the member
is promised one cap for the month:
\begin{equation}\label{eq:commuter-premium}
    \memebership(t, M) = M\,\bar R\rbr{t_0,t_0,K}, \qquad
    \bar R\rbr{t_0,t_0,K} = \frac{1}{\abr{\Pi_h}}\sum_{\path_i \in \Pi_h} R\rbr{t_0,t_0,\path_i,K},
\end{equation}
where $\Pi_h$ collects the sampled history routes. The member pays $\memebership(t, M)$ up front
and then $\min\rbr{P_i, K}$ for each ride $i$ actually taken. The provider absorbs the overrun $\rbr{P_i - K}^{+}$; there is no separate cap on the month's
total. Profit is~\eqref{eq:profit} accumulated over taken rides (30 at max), at cutoff $\zeta = 0$, against the single monthly premium.
Held-out trips are excluded from the travel-time model fit.

We draw $100$ such commuters for each history size $h \in \cbr{2,4,6,8}$, and price under both travel-time models at the plain cap $K = P$ and at
the discounted cap $K = 0.9P$ of Section~\ref{sec:discount-analysis}.

\begin{table}[h!]
    \centering
    \caption{Commuter-membership performance over $100$ counterfactual commuters holding a
    $30$-ride monthly membership on a fixed origin--destination pair at a fixed departure hour,
    where $h$ is the number of past trips on the pair used to price the membership.}
    \begin{tabular}{l l c r r r r c}
    \hline
    Model & Strike & $h$ & \shortstack{Profit\\(\% of fares)} & SE &
    \shortstack{Avg.\ profit\\(\$)} & \shortstack{Max.\ loss\\(\$)} & Win rate \\
    \hline
    Population    & $P$ & 2 & $2.15$ & $0.46$ & $10.86$ & $-86.8$  & $0.77$ \\
                  &     & 4 & $1.71$ & $0.60$ & $9.62$  & $-199.5$ & $0.78$ \\
                  &     & 6 & $2.07$ & $0.53$ & $10.57$ & $-157.2$ & $0.83$ \\
                  &     & 8 & $2.13$ & $0.63$ & $10.91$ & $-238.7$ & $0.79$ \\
    \cline{2-8}
                  & $0.9\,P$ & 2 & $3.57$ & $0.73$ & $18.92$ & $-110.5$ & $0.71$ \\
                  &          & 4 & $2.79$ & $0.83$ & $15.34$ & $-213.7$ & $0.68$ \\
                  &          & 6 & $3.09$ & $0.76$ & $16.17$ & $-177.7$ & $0.73$ \\
                  &          & 8 & $3.94$ & $0.88$ & $22.85$ & $-252.1$ & $0.75$ \\
    \hline
    Trip-specific & $P$ & 2 & $0.00$  & $0.40$ & $-0.54$ & $-130.7$ & $0.65$ \\
                  &     & 4 & $-0.12$ & $0.50$ & $1.84$  & $-193.7$ & $0.67$ \\
                  &     & 6 & $-0.39$ & $0.50$ & $-3.32$ & $-210.5$ & $0.66$ \\
                  &     & 8 & $-0.31$ & $0.59$ & $-1.96$ & $-258.3$ & $0.64$ \\
    \cline{2-8}
                  & $0.9\,P$ & 2 & $0.23$  & $0.57$ & $0.78$  & $-155.2$ & $0.56$ \\
                  &          & 4 & $-0.17$ & $0.63$ & $2.36$  & $-205.4$ & $0.51$ \\
                  &          & 6 & $-0.55$ & $0.60$ & $-4.37$ & $-225.1$ & $0.55$ \\
                  &          & 8 & $-0.08$ & $0.74$ & $0.42$  & $-278.3$ & $0.55$ \\
    \hline
    \end{tabular}
    \label{tab:commuter-metrics}
\end{table}

Table~\ref{tab:commuter-metrics} reports the outcome. Membership seems fairly priced under route conditioning and carries a small margin
without it. At a low monthly cost, all memberships can be profitable. The trip-specific product clears between $-0.6\%$ and $+0.2\%$ of the member's fares
across the whole grid.  All are within one standard error from zero, on a premium of $3.7\%$ of
those fares at $K = P$ and $10.8\%$ at $K = 0.9P$. The population product clears $+1.7\%$ to
$+3.9\%$ on a premium of $5.4\%$ and $12.1\%$ respectively. 

The population model has a larger variance, which adds a buffer to profit margins. The discounted strike is not significantly worse than the
plain cap. The premium absorbs the discount almost exactly, with the mean realized ride price landing between $0.995$ and $1.008$
times the strike at $K = P$ and between $1.105$ and $1.121$ at $K = 0.9P$, against the
$1/0.9 = 1.111$ that the discounted cap implies.
\begin{figure}[h!tbp]
    \centering
    \includegraphics[width=0.85\linewidth]{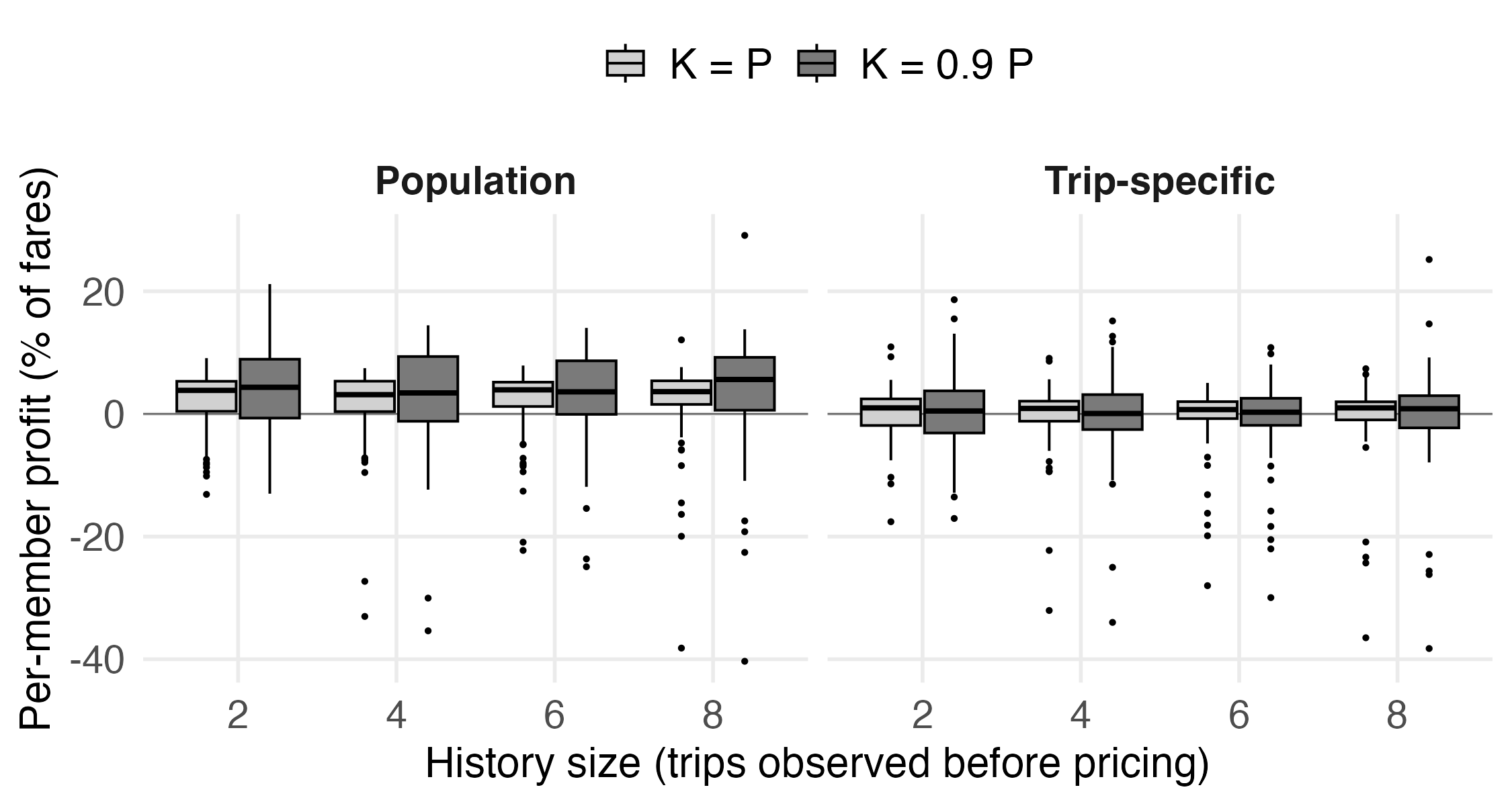}
    \caption{Per-member profit on the commuter membership, as a percentage of the member's fares,
    over $100$ counterfactual commuters, for $K = P$ (light) and $K = 0.9P$ (dark grey).}
    \label{fig:commuter-return}
\end{figure}

Figure~\ref{fig:commuter-return} shows that the margin depends on the number of rides taken. The distribution of profit is tight around zero in both models, and the difference in mean profit is
carried by a left tail of a handful of members, about 2\% of them who overruns the cap.  Accumulated over the month, This exceed the premium collected by $20\%$ to $40\%$ of
their fares. The typical loss-making member is far milder: overruns of some $12\%$ of fares against
a premium of $10\%$, for a net loss of $2$ to $4\%$. The guarantee is therefore doing what it is
priced to do on the typical commuter. 

The dependence on rider history is low. The premium ratio is flat in $h$ ($3.5$--$3.7\%$ for the trip-specific model, $4.8$--$5.0\%$ for the population model).

\begin{figure}[h!tbp]
    \centering
    \includegraphics[width=\linewidth]{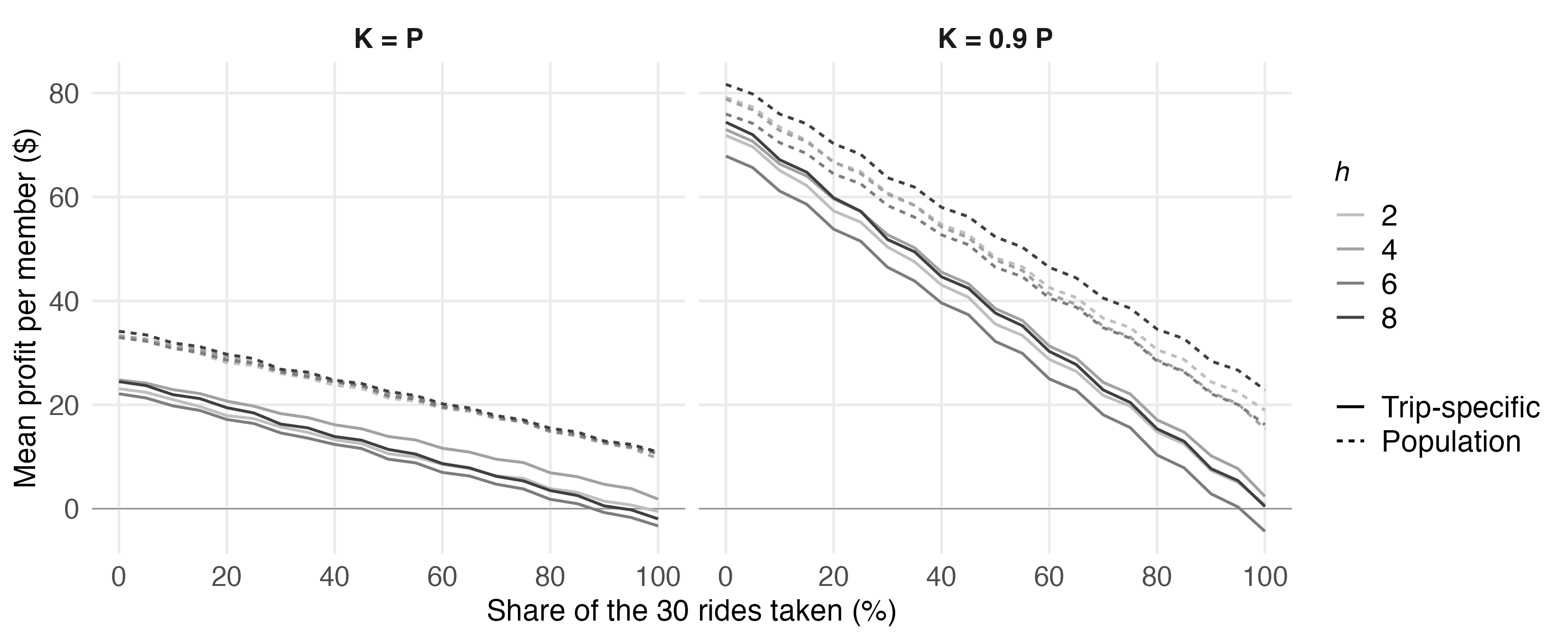}
    \caption{Breakage in the commuter membership: mean profit per member over a
    $30$-ride month as the share of committed rides actually taken rises from $0\%$ to
    $100\%$, at $K = P$ (left) and $K = 0.9P$ (right).}
    \label{fig:commuter-breakage}
\end{figure}

A member pays up front and need not travel every day, and because the premium is collected
regardless, unused rides are pure profit (Fig.~\ref{fig:commuter-breakage}). Averaged over
the grid at $K = P$, the trip-specific return rises from $-0.2\%$ at full use to $+0.7\%$ when
four rides in five are taken and $+2.2\%$ at three in five. The population model follows similarly. 

The exposure is worst against the fully utilizing member (Tab.~\ref{tab:commuter-metrics}), and any realistic attrition moves the product into
profit. Breakage is also the one channel that works in the provider's favour under $K = 0.9P$ the same drop in utilization is worth two to three times as much,
because the premium being kept is itself two to three times larger. Both statements assume the
rides that go unused are dropped at random. Then next section looks at adversarial selection, where the rider exercises selectively.

The commuter product is also where QCD is thinnest. No origin--destination pair in the data
carries more than $13$ trips, and the number of pairs able to supply a commuter falls from
$1{,}001$ at $h=2$ to $63$ at $h=8$, so the larger history sizes draw repeatedly from a small
set of pairs. This is the reason why the history axis is not a direction, but an extrapolation.

\subsection{Selective exercise}\label{sec:selective-exercise}
The rider's private benefit on occasion $i$ is the amount the cap saves them,
$\rbr{P_i - K}^{+}$, which is exactly the provider's payout. Under adverserial selection, the rider will exercise when prices are high. 
Writing $L_i = \rbr{P_i - K}^{+}$ and $A$ for the set of rides taken, the member pays
$\memebership(t,M)$ up front and $\min\rbr{P_i, K}$ on each ride taken, against a cost of
$P_i$ to the provider, so
\begin{equation}\label{eq:selective-profit}
\text{profit} = \memebership(t,M) + \sum_{i \in A} \min\rbr{P_i,K} - \sum_{i \in A} P_i
              = \memebership(t,M) - \sum_{i \in A} L_i .
\end{equation}
Utilization enters only through which $L_i$ are summed. Selection therefore changes the order in which the cost is incurred.

We index the rider's foresight by $\lambda \in \sbr{0,1}$, the rank correlation between the
signal the rider acts on and the realized fare. Let $r_i$ be the rank of $P_i$ among the
month's $M$ fares, $z_i = \Phi^{-1}\rbr{\rbr{r_i - 1/2}/M}$ the corresponding within-month
normal scores, and $\varepsilon_i \sim N(0,1)$ independent of them. The rider exercises in
descending order of
\begin{equation}\label{eq:selective-signal}
S_i\rbr{\lambda} = \lambda\, z_i + \sqrt{1-\lambda^2}\; \varepsilon_i ,
\end{equation}
so the rides taken at utilization $u = k/M$ are the first $k$ of that order and the payout is
$\Lambda\rbr{k,\lambda} = \sum_{i \in A\rbr{k,\lambda}} L_i$. At $\lambda = 0$ the order is
independent of the fare and the design reproduces Section~\ref{sec:commuter} exactly. At
$\lambda = 1$ the rider exercises on the $k$ most expensive mornings of the month. The
ranking is taken on the fare rather than on the payout because $L$ has an atom at zero on the
occasions that finish out of the money, whereas $P$ is continuous and $L$ is monotone in it;
using normal scores rather than the fares themselves makes $\lambda$ scale free. Nothing is
re-priced: the premiums, caps, and realized months are those of Section~\ref{sec:commuter} at
$h = 8$, and the noise $\varepsilon_i$ is drawn once per member and reused at every $\lambda$
and every $k$, so the curves that follow are paired across $\lambda$ member by member.

\begin{figure}[h!tbp]
    \centering
    \includegraphics[width=\linewidth]{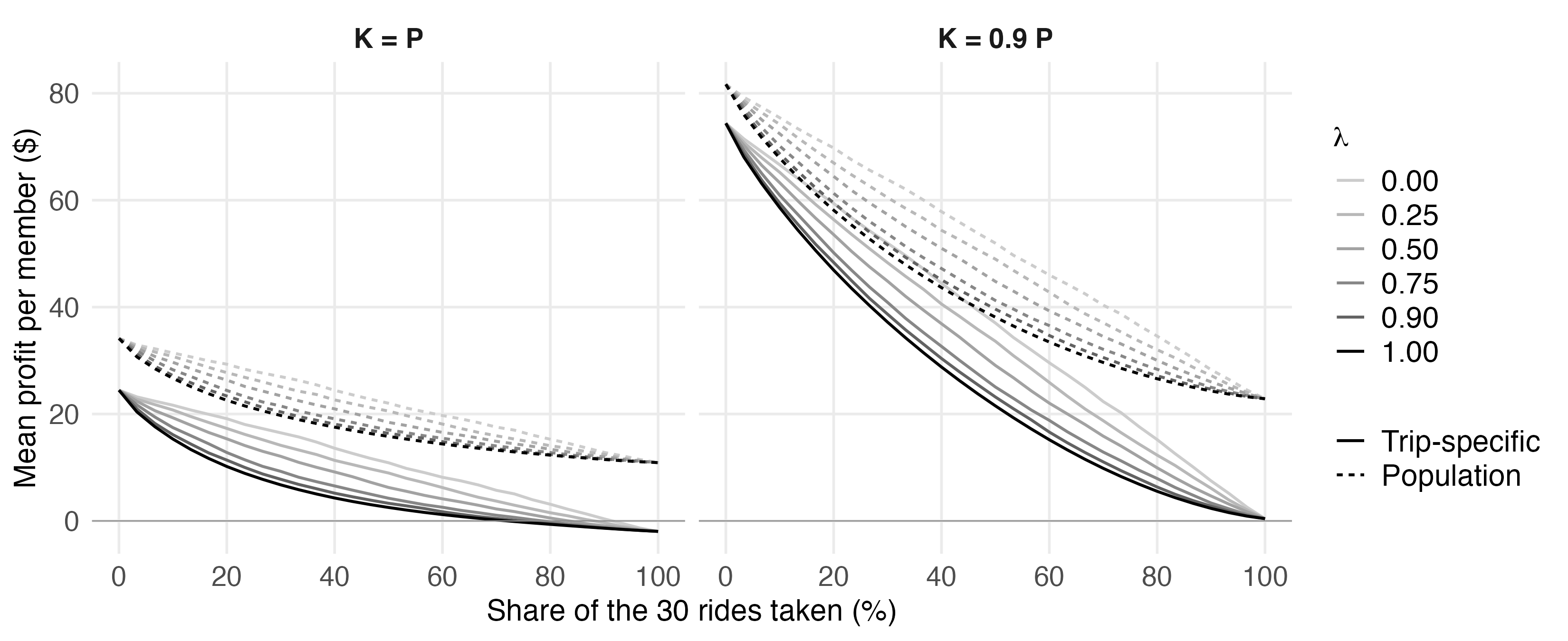}
    \caption{Selective exercise in the commuter membership: mean profit per member over a
    $30$-ride month at $h = 8$, as the share of committed rides taken rises from $0\%$ to
    $100\%$, at $K = P$ (left) and $K = 0.9P$ (right).}
    \label{fig:selective-exercise}
\end{figure}

Figure~\ref{fig:selective-exercise} reports mean profit per member against utilization.
Under random breakage profit falls linearly in $u$, since each dropped ride is worth the
month's average payout. Under selection it falls convexly, the exposure is front-loaded onto
the occasions the rider picks, so the first rides taken are the expensive ones and the last
ones added are nearly free. Every curve meets at $u = 1$, which is
equation~\eqref{eq:selective-profit}. The gap between the linear and convex curves, at
any utilization short of the full month, is the cost of selective exercise.

Table~\ref{tab:selective-exercise} normalizes by the loss ratio $\Lambda/\memebership$ rather
than by a percentage of fares, since the premium is fixed within a member while the fares
taken grow with $k$, and a fares-taken denominator would carry a trend of its own into the
utilization axis. Payouts and premiums are pooled across members, so the ratio crosses one
exactly where mean profit crosses zero.

\begin{table}[ht]
\centering
\caption{Selective exercise at $h = 8$ over the $100$ counterfactual commuters of
Table~\ref{tab:commuter-metrics}, by rider foresight $\lambda$. Upper block: the pooled loss
ratio $\Lambda/\memebership$ at a utilization of $u = 0.8$. Lower block: the break-even
utilization $u^{*}$, the share of the month's rides at which the pooled loss ratio first
reaches one. At full use the loss ratio does not depend on $\lambda$, and equals $0.68$ and
$0.72$ for the population model at $K = P$ and $K = 0.9P$, and $1.08$ and $0.99$ for the
trip-specific model.}
\label{tab:selective-exercise}
\begin{tabular}{l l | r r r r r r}
\hline
Model & Strike & \multicolumn{6}{c}{$\lambda$} \\
\cline{3-8}
 & & $0$ & $0.25$ & $0.50$ & $0.75$ & $0.90$ & $1.00$ \\
\hline
\multicolumn{8}{l}{\textit{Loss ratio at $u = 0.8$}} \\
Population    & $P$      & $0.55$ & $0.59$ & $0.61$ & $0.63$ & $0.64$ & $0.64$ \\
              & $0.9\,P$ & $0.58$ & $0.61$ & $0.63$ & $0.65$ & $0.67$ & $0.67$ \\
Trip-specific & $P$      & $0.87$ & $0.94$ & $0.98$ & $1.01$ & $1.02$ & $1.03$ \\
              & $0.9\,P$ & $0.80$ & $0.84$ & $0.87$ & $0.89$ & $0.92$ & $0.93$ \\
\hline
\multicolumn{8}{l}{\textit{Break-even utilization $u^{*}$}} \\
Population    & $P$      & \multicolumn{6}{c}{$> 1$ at every $\lambda$} \\
              & $0.9\,P$ & \multicolumn{6}{c}{$> 1$ at every $\lambda$} \\
Trip-specific & $P$      & $0.92$ & $0.88$ & $0.83$ & $0.79$ & $0.75$ & $0.72$ \\
              & $0.9\,P$ & \multicolumn{6}{c}{$> 1$ at every $\lambda$} \\
\hline
\end{tabular}
\end{table}

Only the trip-specific product at the plain cap goes negative inside the month. The break-even utilization falls from $u^{*} = 0.92$ under
random breakage to $0.83$ at $\lambda = 0.5$ and $0.72$ under perfect foresight. The population model finishes the month at a loss ratio of $0.68$ and $0.72$, far enough below one. The trip-specific product at the discounted cap lands at $0.994$
at full use, so it sits on the boundary rather than crossing it. This is a direct consequence of the conditioning argument.

With $100$ commuters the \emph{level} of $u^{*}$ is loosely determined. A member-level
bootstrap puts $u^{*}\rbr{0}$ for the trip-specific product at $K=P$ inside
$\sbr{0.63, 0.99}$. The \emph{displacement} is not. Every $\lambda$ is evaluated on the same
members and the same realized months, so the member heterogeneity that dominates the marginal
interval cancels in the contrast, and the paired displacement
$u^{*}\rbr{1} - u^{*}\rbr{0} = -0.20$ carries an interval of $\sbr{-0.33, -0.02}$, entirely
below zero. What this data supports is the direction and the rough size of the movement, not
a point estimate of the threshold itself.

Because no data here identifies a rider's true foresight, the more useful question is: at an assumed utilization $u_0$, how good a forecaster must the rider be before
the product loses money? Writing
$\lambda^{*}\rbr{u_0} = \min\cbr{\lambda : \Lambda\rbr{u_0,\lambda}/\memebership \ge 1}$, the
trip-specific product at $K = P$ gives $\lambda^{*} = 0.69$ at $u_0 = 0.8$ and $0.13$ at
$u_0 = 0.9$, and no attainable $\lambda^{*}$ at $u_0 \le 0.7$. The claim this supports needs
no behavioural calibration. The commuter membership is safe unless the traffic pattern is significantly different from the assumed one, such that the rider is able to consistently 
take advantage of price miscalibration. With adequate traffic model, the product is breakeven under full utilization. 

\section{Discussion} \label{sec:discussion}
 
Providers of on-demand transportation build increasingly complex models to
find a sweet spot between profitability and competitiveness. These models
typically rely on point estimates of travel time. Integrating uncertainty into the pricing model has proven elusive
for two reasons: analytical pricing tools for it have been lacking, and the
travel-time distribution itself is difficult to characterize on a real network.
This work supplies both pieces. Taking the asymptotic travel-time law of~\cite{elmasri2020predictive} as an arithmetic Brownian motion on a metric graph, we
priced the right to a capped fare as a contingent claim, obtaining a
closed-form premium~\eqref{eq:european-knock-in} under both a
population and a route-conditional travel-time model.
 
Empirically, the guarantee is inexpensive. On QCD the premium runs between
$0.4\%$ and $4.4\%$ of the fare across strikes, and conditioning on the realized
route roughly halves it. For example, $2.15\%$ versus $4.43\%$ at the
expected-price strike, while also shrinking the worst-case single-ride loss
(near \$22 versus \$29). The route-conditional model is therefore the better
product for the on-demand guarantee: tighter variance yields a cheaper premium
with a thinner tail. That same tightness reverses once a flat discount is
layered on: the discounted trip-specific membership loses its buffer against
overruns and turns marginally loss-making, while the coarser population model
retains a small margin. The lesson is that the route conditioning which makes
the bare guarantee cheap also removes the slack a bundled discount needs, so the
two design choices interact and should not be tuned in isolation.
 
A unifying way to read the membership results is that abuse exploits precisely
the dimension along which the price fails to condition. The commuter
membership fixes the origin--destination pair and the departure hour, and thereby
conditions on both length and timing; nothing but travel-time variance on a known
commute is left for the holder to select on. Priced that way it is close to fair
rather than exposed: over the grid of Table~\ref{tab:commuter-metrics} the
route-conditional product returns between $-0.6\%$ and $+0.2\%$ on a premium of
$3.7\%$ of the member's fares. A shortfall in utilization moves it into profit
only when the rides forgone are uninformative. Section~\ref{sec:selective-exercise} stress-tests the residual channel and finds
that a holder who can rank their own travel times pulls the break-even
utilization down from $92\%$ to $72\%$, so the exposure is worst against the
selectively utilizing member rather than the fully utilizing one. Per-ride pricing conditions on everything, leaving no such channel at all. Read together, the abuse grid and these three products trace a single axis---the
more the price conditions, the less room a rational holder has to select against it.

For example, the unrestricted flat membership conditions on nothing: a single strike $K$ covers any route at any
time, so the rider's marginal price of distance is zero and a rational holder
substitutes toward the longest covered trips, with the provider's per-ride loss
approaching $d_0 - K$, the full price gap on a long ride.
 
This logic suggests a class of products that are abuse-proof by construction. Those whose subsidy is additive in the fare rather than attached to its variable
part. Consider a bundle granting a fixed \$$D$ discount on each of $N$ rides.
Because the discount is a constant, it leaves the rider's marginal price of
distance and time unchanged; there is no incentive to lengthen or retime trips.
The provider's liability is then deterministic and bounded,
\begin{equation}
  \mathrm{Cost} \;=\; \sum_{i=1}^{N} \min(D, P_i)
  \;=\; ND - \sum_{i=1}^{N}(D - P_i)^{+} \;\le\; ND,
  \label{eq:additive-cost}
\end{equation}
with expected value $\mathbb{E}[\mathrm{Cost}] = ND - N\,\mathbb{E}[(D-P)^{+}]$.
The only stochastic term is the correction for rides whose fare falls below $D$,
on which the non-negativity of price caps the rebate at the fare itself. No travel-time model is required to price the product, only the fraction of fares
below $D$. The maximum liability is known in advance. Moving the subsidy
from the level of the fare (a capped total, marginal length-price zero) to an
additive per-ride amount (marginal length-price preserved) converts an
uninsurable moral-hazard problem into a bounded, closed-form cost. Product
design, not pricing sophistication, is what neutralises the abuse.
 
These conclusions inherit the assumptions under which they were derived, and
their limits mark the boundary of the claims. Our pricing is deliberately
cost-based and set in a monopoly market with immediate fulfilment (A1--A3): we
price the guarantee written on top of the fare process rather than the
market-clearing fare itself, and we exclude the matching, strategic-driver, and
supply--demand-equilibrium mechanisms that a two-sided model would carry. Riders
are taken to be price-insensitive beyond the modelled price (A4), so the abuse
analysis captures selection on route and timing but not elastic demand response;
a rider who takes more or fewer trips because the membership exists is outside
the present model. The frictionless assumptions (B1--B2) make the discounting exact but abstract from
real settlement costs. Finally, because road travel time is not a traded,
hedgeable asset, the market is incomplete and no replicating portfolio exists;
our premium is a discounted expected cost under the modelled law, not a
no-arbitrage replication price, and it is therefore only as good as that law's
fit. The out-of-sample validation of Section~\ref{sec:fedelity-analysis} supports this claim. 
 
Several extensions follow naturally. Relaxing price-insensitivity (A4) would let
the framework speak to demand response and to the joint design of price and
membership, at the cost of the two-sided machinery we set aside. Incorporating a
non-zero and stochastic interest rate (B1) would matter for genuinely long-dated
products. Scheduled rides booked months ahead, or memberships priced a year
out, where the value-of-money adjustment stops being negligible. On the data
side, the samplers of Section~\ref{sec:sampler} make the method portable to any
network with edge-level travel-time statistics, so replicating the QCD study on
a second city would test how far the premium magnitudes and the abuse geometry
generalize. Most promising, the additive-subsidy result points toward a broader
design question: given a target subsidy and a tolerance for selection, which
contract shapes are both attractive to riders and bounded for the provider? The
guarantee, the capped membership, and the fixed per-ride discount are three
points on that spectrum; characterizing it fully is left to future work.

\section*{Funding}
Mohamad Elmasri acknowledges financial support from the Natural Sciences and Engineering Research Council of Canada (NSERC) [
RGPIN-2025-04956, DGECR-2025-0022]. Yunran Wei acknowledges financial support from NSERC [RGPIN-2023-04674, DGECR-2023-00454]. Both acknowledge the start-up fund at Carleton University.
\begin{appendices}

\section{Derivation of membership and scheduled rides}\label{app:derivations}

This appendix carries out the intermediate steps omitted
from~\eqref{eq:european-knock-in},~\eqref{eq:scheduled-price}
and~\eqref{eq:membership-price}. Throughout, $\T_\path$ is the travel time along
$\path$, the constants $d_0,\dots,d_4$ are those defined below
\eqref{eq:european-knock-in}, and expectations are taken under the travel-time
law of Section~\ref{sec:traveltime-process}.

\subsection*{On-demand guarantee~\eqref{eq:european-knock-in}}

Substituting the liability~\eqref{eq:ride-cost} into the discounted expectation
and discounting over the ride itself,
\begin{equation}\label{eq:app-knock-in}
\begin{aligned}
 R(t,t_0, \rho, K) &= \EE\sbr{e^{-r(t_0-t)}R(t_1,t_0, \path, K)} \\
  &= \EE\sbr{e^{-r(t_0-t) -r\T_\path}\rbr{P_\path(\pend, t_1) - K}^+ \one_{\cbr{P_\path(\pend, t_1) \geq Ke^\zeta}}}  \\
  &=e^{-r(t_0-t)} e^{-rd_4}\sbr{\cbr{d_0 - rC_1d_1^2- K}\cbr{1-\Phi\rbr{d_3}} + C_1d_1 \phi(d_3)}.
\end{aligned}
\end{equation}
The second line uses $t_1 = t_0 + \T_\path$, so that the discount factor from
request to arrival splits as $e^{-r(t_0-t)}e^{-r\T_\path}$. The third line
evaluates the Gaussian expectation: the realized price
$P_\path(\pend,t_1) = C_0 + C_1\T_\path + C_2\delta_\path$ is affine in
$\T_\path$, so its mean is $d_0$ and its standard deviation is $C_1d_1$, and the
knock-in event $\cbr{P_\path(\pend,t_1) \geq Ke^\zeta}$ is the event that a
standard normal exceeds $d_2$. The factor $e^{-r\T_\path}$ tilts the
distribution, which shifts the threshold from $d_2$ to $d_3 = d_2 + rd_1$,
contributes the deterministic factor $e^{-rd_4}$, and lowers the mean by
$rC_1d_1^2$.

\subsection*{Scheduled ride~\eqref{eq:scheduled-price}}

The route is unknown at request time, so the guarantee is averaged over
$\Pi\rbr{\pstart, \pend}$ with weights $\pi(\path_i)$,
\begin{equation}\label{eq:app-scheduled}
\begin{aligned}
    \scheduled(t, t_0, K) & = 
\sum_{\path_i \in \Pi\rbr{\pstart, \pend}} R(t, t_0, \path_i, K)\pi(\path_i) \\
& = \sum_{\path_i \in \Pi\rbr{\pstart, \pend}} \EE\sbr{e^{-r(t_0-t + \T_{\path_i})}R(t_1, t_0, \path, K)}\pi(\path_i) \\
 & = \sum_{\path_i \in \Pi\rbr{\pstart, \pend}} \EE\sbr{\EE \sbr{e^{-r(t_0-t) - r\T_{\path_i}}R(t_1, t_0, \path, K)\given t_0}}\pi(\path_i) \\
& = e^{-r(t_0-t)} \sum_{\path_i \in \Pi\rbr{\pstart, \pend}} \EE\sbr{e^{-r\T_{\path_i}}R(t_1, t_0, \path, K)}\pi(\path_i).
\end{aligned}
\end{equation}
The second line inserts the first line of~\eqref{eq:app-knock-in} for each
route. The third conditions on the pickup time $t_0$ and applies the tower
property. The fourth uses that $t_0$ is known at request time for a scheduled
ride, so $e^{-r(t_0-t)}$ is deterministic and factors out of both the
expectation and the sum.

\subsection*{Membership~\eqref{eq:membership-price}}

A membership fixes neither the route nor the pickup time, so the guarantee is
averaged over both, with $M$ rides priced identically,
\begin{equation}\label{eq:app-membership}
\begin{aligned}
    \memebership(t, M) & = M
    \sum_{t_0 \in [0, T_R]} 
    \sum_{\path_i \in \Pi\rbr{\pstart, \pend}} 
    R\rbr{t_0, t_0, \path_i, K_{\path_i}}
    \pi(t_0) \pi(\path_i), \\
    & = M \sum_{t_0 \in T_R} e^{-r\rbr{t_0 - t}}
    \sum_{\path_i \in \Pi\rbr{\pstart, \pend}} 
    R\rbr{t_0, t_0, \path_i, K}
    \pi(t_0) \pi(\path_i), \\
      & = M \sum_{t_0 \in  T_R} e^{-r\rbr{t_0 - t}}
    \sum_{\path_i \in \Pi\rbr{\pstart, \pend}} 
    \EE\sbr{e^{-r\T_{\path_i}}R\rbr{ t_1, t_0, \path_i, K}}
    \pi(t_0) \pi(\path_i), \\
    & = M \sum_{t_0 \in T_R} e^{-r\rbr{t_0 - t}} \scheduled\rbr{t_0, t_0, K}   \pi(t_0).
\end{aligned}
\end{equation}
The second line fixes a common strike $K$ across routes and pulls out the
discount from purchase to pickup. The third replaces each per-route guarantee by
the expectation in the last line of~\eqref{eq:app-knock-in}. The fourth
recognizes the inner route average as the scheduled
price~\eqref{eq:app-scheduled} evaluated at $t = t_0$, so that a membership is a
$\pi(t_0)$-average of $M$ scheduled rides.

\end{appendices}

\bibliographystyle{plainnat}
\bibliography{references}

\end{document}